\documentclass[12pt]{article}
\usepackage{graphicx}
\begin{document}
\newcommand{\beq}{\begin{equation}}
\newcommand{\eeq}{\end{equation}}
\newcommand{\beqa}{\begin{eqnarray}}
\newcommand{\eeqa}{\end{eqnarray}}
\newcommand{\beqar}{\begin{eqnarray*}}
\newcommand{\eeqar}{\end{eqnarray*}}
\newcommand{\al}{\alpha}
\newcommand{\be}{\beta}
\newcommand{\del}{\delta}
\newcommand{\D}{\Delta}
\newcommand{\eps}{\epsilon}
\newcommand{\ga}{\gamma}
\newcommand{\Ga}{\Gamma}
\newcommand{\ka}{\kappa}
\newcommand{\nn}{\nonumber}
\newcommand{\inn}{\!\cdot\!}
\newcommand{\h}{\eta}
\newcommand{\ii}{\iota}
\newcommand{\kk}{\varphi}
\newcommand\F{{}_3F_2}
\newcommand{\la}{\lambda}
\newcommand{\La}{\Lambda}
\newcommand{\na}{\prt}
\newcommand{\Om}{\Omega}
\newcommand{\om}{\omega}
\newcommand{\p}{\phi}
\newcommand{\sig}{\sigma}
\renewcommand{\t}{\theta}
\newcommand{\z}{\zeta}
\newcommand{\ssc}{\scriptscriptstyle}
\newcommand{\eg}{{\it e.g.,}\ }
\newcommand{\ie}{{\it i.e.,}\ }
\newcommand{\labell}[1]{\label{#1}} 
\newcommand{\reef}[1]{(\ref{#1})}
\newcommand\prt{\partial}
\newcommand\veps{\varepsilon}
\newcommand{\pol}{\varepsilon}
\newcommand\vp{\varphi}
\newcommand\ls{\ell_s}
\newcommand\cF{{\cal F}}
\newcommand\cA{{\cal A}}
\newcommand\cS{{\cal S}}
\newcommand\cT{{\cal T}}
\newcommand\cV{{\cal V}}
\newcommand\cL{{\cal L}}
\newcommand\cM{{\cal M}}
\newcommand\cN{{\cal N}}
\newcommand\cG{{\cal G}}
\newcommand\cH{{\cal H}}
\newcommand\cI{{\cal I}}
\newcommand\cJ{{\cal J}}
\newcommand\cl{{\iota}}
\newcommand\cP{{\cal P}}
\newcommand\cQ{{\cal Q}}
\newcommand\cg{{\it g}}
\newcommand\cR{{\cal R}}
\newcommand\cB{{\cal B}}
\newcommand\cO{{\cal O}}
\newcommand\tcO{{\tilde {{\cal O}}}}
\newcommand\bz{\bar{z}}
\newcommand\bc{\bar{c}}
\newcommand\bw{\bar{w}}
\newcommand\bX{\bar{X}}
\newcommand\bK{\bar{K}}
\newcommand\bA{\bar{A}}
\newcommand\bZ{\bar{Z}}
\newcommand\bxi{\bar{\xi}}
\newcommand\bphi{\bar{\phi}}
\newcommand\bpsi{\bar{\psi}}
\newcommand\bprt{\bar{\prt}}
\newcommand\bet{\bar{\eta}}
\newcommand\btau{\bar{\tau}}
\newcommand\hF{\hat{F}}
\newcommand\hA{\hat{A}}
\newcommand\hT{\hat{T}}
\newcommand\htau{\hat{\tau}}
\newcommand\hD{\hat{D}}
\newcommand\hf{\hat{f}}
\newcommand\hg{\hat{g}}
\newcommand\hp{\hat{\phi}}
\newcommand\hi{\hat{i}}
\newcommand\ha{\hat{a}}
\newcommand\hb{\hat{b}}
\newcommand\hQ{\hat{Q}}
\newcommand\hP{\hat{\Phi}}
\newcommand\hS{\hat{S}}
\newcommand\hX{\hat{X}}
\newcommand\tL{\tilde{\cal L}}
\newcommand\hL{\hat{\cal L}}
\newcommand\tG{{\widetilde G}}
\newcommand\tg{{\widetilde g}}
\newcommand\tphi{{\widetilde \phi}}
\newcommand\tPhi{{\widetilde \Phi}}
\newcommand\te{{\tilde e}}
\newcommand\tk{{\tilde k}}
\newcommand\tf{{\tilde f}}
\newcommand\ta{{\tilde a}}
\newcommand\tb{{\tilde b}}
\newcommand\tR{{\tilde R}}
\newcommand\teta{{\tilde \eta}}
\newcommand\tF{{\widetilde F}}
\newcommand\tK{{\widetilde K}}
\newcommand\tE{{\widetilde E}}
\newcommand\tpsi{{\tilde \psi}}
\newcommand\tX{{\widetilde X}}
\newcommand\tD{{\widetilde D}}
\newcommand\tO{{\widetilde O}}
\newcommand\tS{{\tilde S}}
\newcommand\tB{{\widetilde B}}
\newcommand\tA{{\widetilde A}}
\newcommand\tT{{\widetilde T}}
\newcommand\tC{{\widetilde C}}
\newcommand\tV{{\widetilde V}}
\newcommand\thF{{\widetilde {\hat {F}}}}
\newcommand\Tr{{\rm Tr}}
\newcommand\tr{{\rm tr}}
\newcommand\STr{{\rm STr}}
\newcommand\hR{\hat{R}}
\newcommand\M[2]{M^{#1}{}_{#2}}
 
\begin{titlepage}
\begin{center}

\vskip 2 cm
{\LARGE \bf  Duality constraints on  effective actions  
 }\\
\vskip 1.25 cm
  Mohammad R. Garousi\footnote{garousi@um.ac.ir}

\vskip 1 cm
{{\it Department of Physics, Ferdowsi University of Mashhad\\}{\it P.O. Box 1436, Mashhad, Iran}\\}
\vskip .1 cm
 \end{center}

\begin{abstract}
Superstring theories at low energy limit are described by the corresponding supergravities, and their non-perturbative D-brane/O-plane excitations   are described by DBI and WZ   actions. Higher derivative corrections to these effective actions are important for  understanding the stringy behavior of the fundamental objects. They   may be extracted from the contact terms of the corresponding   S-matrix elements.  On the other hand, the superstring theories enjoy the  T- and S-dualities which appear in the S-matrix elements as duality Ward identities. These Ward identities might   be used as generating functions for constructing the S-matrix elements.  The dualities may also be used directly to construct the   effective actions. In this article, we   review the duality Ward identities which can be used to generate S-matrix elements, and review the  dualities which may be used directly to construct the higher derivative corrections to the effective actions. 
\end{abstract}
\end{titlepage}
\tableofcontents

\section{Introduction   } \label{intro}

In the first revolution of string theory, it has been cleared that there are only  five different anomaly free superstring theories, \ie type I, type IIA, type IIB, $SO(32)$ and $E_8\times E_8$ Heterotic theories, which live in ten dimensional spacetime \cite{Green:1984sg,Green:1984qs,Harvey:2005it}. In the second revolution, the study of string dualities reveals that some of the superstring theories have  extended non-perturbative objects like D$_p$-branes and O$_p$-planes \cite{Polchinski:1995mt,Polchinski:1996na}. It has been also found that the five superstring theories are interconnected and are  some faces of the eleven dimensional M-theory \cite{Schwarz:1996bh}. An important tool for exploring these theories  and their non-perturbative objects  is the    effective action.  The effective actions of the superstring theories at low energy are given by the corresponding supergravities \cite{Howe:1983sra,Witten:1995ex} and the effective action of D$_p$-branes/O$_p$-planes  are given by Diract-Born-Infeld (DBI) and Wess-Zumino (WZ) actions \cite{Bachas:1995kx,Douglas:1995bn}. Even though these effective actions are enough for exploring many aspects of string theory, \eg AdS/CFT duality \cite{Maldacena:1997re}, there are important situations that one needs to find the higher derivative corrections to these effective actions \eg  to explore the string landscape \cite{Susskind:2003kw}.

For example, let us review the  compactification of the 10-dimensional type IIB superstring theory  to  maximally symmetric 4-dimensional spacetime (see \eg \cite{Becker:2007zj}). At low energy, the theory is described by the  type IIB supergravity which is
\beqa
S_{IIB}&\!\!\supset\!\!&\frac{1}{2\kappa^2}\int d^{10}x\sqrt{-G}[R-\frac{|\prt\tau|}{2({\rm Im}\,\tau)^2}-\frac{|G_3|}{2({\rm Im}\,\tau)^2}-\frac{|F_5|^2}{4}]+\frac{1}{8i\kappa^2}\int \frac{C_4\wedge G_3\wedge G_3^*}{{\rm Im}\,\tau}\labell{sugraB}
\eeqa
where the metric is  in the Einstein frame,  $\tau=C_0+ie^{-\phi}$, the RR five-form field strength is $F_5=dC_4-(C_2\wedge dB-B\wedge dC_2)/2$, which is constrained to  satisfy the  self-duality condition $F_5=\star_{10}F_5$ at the  level of the equations of motion  \cite{Bergshoeff:1995sq}, and $G_3=dC_2-\tau\,dB$.   If the 10-dimensional spacetime is product of the 4-dimensional spacetime  and a 6-dimensional internal compact manifold and if there is  no  flux  in the 6-dimensional   manifold, \ie
\beqa
ds^2_{10}&=& \eta_{\mu\nu}dx^{\mu}dx^{\nu}+ g_{mn}(y)dy^{\mu}dy^{\nu}\nn\\
F_5&=&G_3=\prt\tau=0 
\eeqa
 one would find many 4-dimensional massless scalars. None of which, however,  appears in our real world! This is called moduli space problem.

To solve this problem, one may consider more general setting of warped compactification with flux,\ie
\beqa
ds^2_{10}&=&e^{2A(y)}\eta_{\mu\nu}dx^{\mu}dx^{\nu}+e^{-2A(y)}g_{mn}(y)dy^{\mu}dy^{\nu}\nn\\
F_5&=&(1+\star_{10})d\alpha(y)\wedge dx^0\wedge dx^1\wedge dx^2\wedge dx^3\nn\\
G_3&=&G_{mnp}(y)dx^m\wedge dx^n\wedge dx^p\nn\\
\tau&=&\tau(y)
\eeqa
where $A(y)$ is the warped factor. With the above anzats,  the supergravity equations of motion produce the following tadpole equation:
\beqa
 \nabla^2 e^{4A}&=&\frac{e^{8A}}{2{\rm Im}\,\tau}|G_3|^2+e^{-4A}(|\prt\alpha|^2+|\prt e^{4A}|^2)\labell{tad}
\eeqa
where $\nabla^2$ is the Laplacian in the internal manifold. Since all terms on the right hand side are positive, one finds $G_3=F_5=A=0$ upon    integrating over the internal manifold.  In other words, if one considers the effective action of  the type IIB superstring theory at the leading order of $\alpha'$, then one would find   that it is impossible to compact type IIB superstring theory on a warped manifold or on  a manifold with fluxes. 

One may also consider internal manifolds with $(p-3)$-cycles $\Sigma$ on which the non-perturbative objects D$_p$-branes or O$_p$-planes are wrapped. The energy momentum tensor of the branes 
 appear on the right hand side of the tadpole equation \reef{tad}. If one describes the D$_p$-brane or O$_p$-plane effectively by the DBI and WZ actions at the leading order of $\alpha'$, 
 then one would find the WZ part produces zero energy momentum tensor, and the DBI part produces positive contribution to the right hand side of the tadpole equation \reef{tad}. As a result, the tadpole equation again does not allow to have internal manifolds in which   branes are wrapped   on its cycles. In other words, the moduli space problem could not be solved with the effective actions at the leading order of $\alpha'$.   

How the $\alpha'$-corrections to the effective actions can solve the problem? One particular set of higher derivative correction to the WZ part at order $\alpha'^2$  which has been found from anomaly cancellation mechanism \cite{Green:1996dd,Cheung:1997az,Minasian:1997mm}, is the following:
\beqa
S_{p}\supset\frac{\pi^2\alpha'^2}{24}\mu_p\int_{R^4\times\Sigma}C_{p-3}\wedge(tr R_T\wedge R_T-trR_N\wedge R_N)\labell{4-der}
\eeqa
The energy-momentum tensor of this term produces a negative contribution to the right hand side of the tadpole equation \reef{tad}. If one includes only this     $\alpha'^2$-correction    to the effective actions of the branes, then the warped compactification to the internal manifold with branes and fluxes are allowed. Moreover, if one includes one particular non-perturbative effect, then all four-dimensional scalar fields are constrained by  some potentials and the moduli space problem would be solved \cite{Kachru:2003aw}. However, there are many discrete vacua for the  potentials which produce the string landscape \cite{Susskind:2003kw}.  

Higher derivative corrections to the supergravities are also important for finding the discrete vacua. Consider, for example, the  compactification of the 11-dimensional M-theory   to  the maximally symmetric 3-dimensional spacetime (see \eg \cite{Becker:2007zj}). At low energy, the M-theory is described by the  11-dimensional supergravity which is 
\beqa
S_{\rm M}& \supset&\frac{2\pi}{(2\pi\ell_p)^9}\int d^{11}x \sqrt{-g}(R-\frac{1}{2}|F_4|^2)-\frac{ \pi}{3(2\pi\ell_p)^9}\int A_3\wedge F_4\wedge F_4\labell{2-der}
\eeqa
Consider the warped compatification into 8-dimensional internal manifold with flux, \ie
\beqa
ds^2_{11}&=&e^{-A(y)}\eta_{\mu\nu}dx^{\mu}dx^{\nu}+e^{A(y)/2}g_{mn}(y)dy^{\mu}dy^{\nu}\nn\\
F_4&=&F_{mnpq}(y)dx^m\wedge dx^n\wedge dx^p\wedge dx^q 
\eeqa 
The 4-form equation of motion gives the following tadpole equation:
\beqa
 \nabla^2 A(y)&=&-|F_4|^2\labell{tad2}
\eeqa
After integration over the internal manifold, one again finds $F_4=0$ and the warp factor $A(y)$ is a constant. Therefore,  there would be  no potential for the 3-dimensional scalars, \ie there would be the moduli space problem, if one describes the M-theory by the 2-derivative effective action \reef{2-der}. The moduli space problem in this case may be solved by including the following    8-derivative correction to the 11-dimensional supergravity \cite{Duff:1995wd}, \ie
\beqa
S_{M}&\supset&-\frac{2\pi}{(2\pi\ell_p)^3}\int A_3\wedge X_8\labell{8-der}
\eeqa
where the 8-form has been found from anomaly cancellation mechanism to be 
\beqa
X_8&=&\frac{1}{(2\pi)^4}[\frac{1}{192}tr R\wedge R\wedge R\wedge R-\frac{1}{768}(R\wedge R)^2]
\eeqa
The above higher derivative term has contribution $-\frac{1}{2}(2\pi\ell_p)^6X_8$ to the right hand side of the tadpole equation \reef{tad2}. After integration over the internal manifold, one finds 
\beqa
\int_M |F_4|^2&=&-\frac{1}{2}(2\pi\ell_p)^6\int_M X_8\,=\,\frac{1}{48}(2\pi\ell_p)^6  \chi
\eeqa
where $\chi$ is the Euler character of the internal manifold. Therefore, the   8-derivative term \reef{8-der} makes the warped compactification with flux to be possible. The fluxes and non-perturbative effects, on the other hand, produce potential for the 3-dimensional scalar fields which solves the moduli space problem and produces an M-theory  landscape.

The  curvatures corrections  \reef{4-der} and \reef{8-der}    are found from anomaly cancellation mechanism \cite{Green:1996dd,Cheung:1997az,Minasian:1997mm,Duff:1995wd}. However, as we will review in section 3, the metric transforms to B-field, to dilaton and to all RR forms under the sequences of the T-duality and S-duality transformations, so there must be  many other 4-derivative corrections to the brane action  and many other 8-derivative corrections to the supergravities  which can not be found from anomaly cancellation mechanism and  may have effects in finding the true vacua. We are interested in these couplings.

There are different approaches for constructing the higher derivative effective actions in the string theory.  One is the non-linear sigma model   which constrains the two-dimensional world-sheet   theory in the presence of background fields to be conformal invariant \cite{Callan:1985ia,Tseytlin:1988rr}. We are not  interested in this approach in this review article. Another approach for finding such higher derivative terms is   the S-matrix approach \cite{Green:1981xx,Green:1981ya} which we will review in the next section. This method is appropriate for calculating the higher derivatives of metric  because derivatives of metric appear covariantly in the  curvature which has two derivatives. As a result, to find the 8-derivative corrections to the supergravity, one needs to calculate the sphere-level  S-matrix element of four graviton vertex operators,  and   to find the 4-derivative corrections to the brane action, one needs to calculate the disk-level or $PR^2$-level  S-matrix element of two graviton vertex operators. However, since there is a conservation of momentum in the  S-matrix elements,   this calculation fixes neither    the four-curvature couplings in the supergravity which are total derivatives at four-graviton level, nor the two-curvature couplings in the brane action which are world-volume total derivatives at   two-graviton level. They may be fixed by other methods or by studying the higher-point functions. Unlike the supergravity and the DBI/WZ actions which have neither genus nor non-perturbative correction, their higher derivative corrections are not complete unless one includes their corresponding genus and non-perturbative corrections. The genus corrections can be extracted from the corresponding loop-level S-matrix elements. The genus and the non-perturbative corrections  in type IIB theory  may   be found from requiring the tree-level couplings to be consistent with S-duality \cite{Green:1997tv,Bachas:1999um}. 

The higher derivatives of  other NSNS fields or   RR fields appear in their corresponding field strengths. As a result, the 8-derivative corrections to the supergravities require    the sphere-level S-matrix element of five, six, seven  and eight vertex operators,  and   the 4-derivative corrections to the brane action require the disk-level/$PR^2$-level S-matrix element of three and four vertex operators. Such calculations are technically very complicated. So one has to use other methods for finding such tree-level higher derivative corrections. Supersymmetry, in which we are not interested in this review,  may be able to find all such couplings \cite{Howe:1983sra,Nilsson:1986rh,Paban:1998ea,Green:1998by} including  the moduli-dependnence of the type IIB theory \cite{Green:1998by}. String duality may also be able to find these couplings.  We will review in section 3 the duality method which enables one to find all  couplings at each order of $\alpha'$ by requiring the tree-level gravity couplings to be consistent with T-duality and S-duality \cite{Garousi:2009dj,Becker:2010ij,Garousi:2010rn,Garousi:2011fc,Garousi:2012yr,Garousi:2013nfw,Garousi:2011fc,Garousi:2012yr,Garousi:2013nfw,Liu:2013dna,Robbins:2014ara,Garousi:2014oya}.  The idea that T- and S-dualities put constraints in  the   effective string actions at the leading order of $\alpha'$  appeared at the first time 
in \cite{Ferrara:1989bc,Font:1990gx}.

An outline of the review is as follows: In section 2, we briefly review the Polyakov prescription for constructing the S-matrix elements in perturbative string theory. In subsection 2.1, we explicitly calculate the sphere-level S-matrix element of four closed string tachyons in bosonic string theory which reproduces the Virasoro-Shapiro amplitude. In subsection 2.2, we review the calculation of  the sphere-level S-matrix element of four NSNS vertex operators   in type II superstring theory. We demonstrate how this amplitude at low energy produces 8-derivative  corrections to the type II supergravity involving the Riemann curvature and the second covariant derivative of dilaton and B-field. In subsection 2.3, we review the calculation of the disk-level S-matrix element of two NSNS or RR vertex operators in type II supestring theory and show how the low energy limit of this amplitude produces 4-derivative corrections to the DBI and WZ action involving the Riemann curvature and the second covariant derivative of dilaton, B-field and RR-forms. In
subsection 2.4, we repeat the calculation of the previous subsection for projective plane instead of disk. 

In section 3, we review the well-known dualities of the string theory. In subsection 3.1, we  briefly review the T-duality of the spectrum of the bosonic string theory when compactified on a tours $T^n$, to extract the T-duality transformations of the scalar fields that parametrize the tours. We then use the path-integral method   to extend these transformations to the curved spacetime with background fields to find the Buscher rules. The DBI action is invariant under the Buscher rules. We then use the  constraint that the  WZ action must also be invariant under the T-duality transformation, to rederive the standard T-duality transformation of the RR fields.   Using the observation that the effective actions at the leading order of $\alpha'$ are invariant under the T-duality transformation, one expects that the covariant higher derivative corrections to the effective actions to be also invariant under the T-duality transformations. However, the T-duality transformations  are modified by the covariant higher derivative corrections. Alternatively,   the  invariance of effective actions  under the standard T-duality transformations requires the higher derivative terms to be non-covariant. A non-covariant  field redefinition may change the non-covariant higher derivative couplings to the covariant couplings. 

In subection 3.2, we review the S-duality and in particular the $SL(2,R)$ transformations of the massless fields that appear in the type II effective actions and show how the invariance of the effective actions under these transformations may fix the genus and the non-perturbative corrections to the effective actions. We show how the S-duality  may be used to find new tree-level couplings by imposing the couplings found in section 2 to be invariant under the $SL(2,R)$ transformations. 
In section 4, we review the observation that the S-matrix elements must satisfy the duality Ward identity. We demonstrate how these Ward identities may be used to generate new S-matrix element from   a given S-matrix element, and review the works that have been done in support of this observation.  

In section 5, we review  the  specific example of the O$_p$-plane effective action in type II superstring theory that the T-duality constraint is used to find all NSNS 4-derivative  corrections to this action. The higher derivative couplings are covariant and the T-duality transformations are also the Buscher rules. In this case, we know from the supergravity corrections that the T-duality does not receive higher derivative corrections at order $\alpha'^2$, as a result the covariant action at order $\alpha'^2$ is consistent with the Buscher rule.  In section 6, we review the specific example of O$_p$-plane/D$_p$-brane effective action in the bosonic string theory. The T-duality constraint is used to find the covariant O$_p$-plane effective action completely   at order $\alpha'$. In this case, however, the T-duality transformation is the Buscher rule plus its $\alpha'$-correction. The above T-duality  constraint has been also used to find the covariant D$_p$-brane effective action at order $\alpha'$ for only massless closed string fields up to terms that contains   B-field potential. In this section, we also review the construction of a non-covariant D$_p$-brane effective action at order $\alpha'$ which includes only  masslesss open string fields to all orders. The T-duality transformation that has been used is the standard T-duality transformation for the massless open string fields without the $\alpha'$-corrections. In section 7, we briefly discuss the new calculations that may be done by the duality method that we have reviewed here.

\section{S-matrix elements in perturbative string theory} 

In quantum field theory with specific spacetime action $S[\Phi]=S_{\rm free}[\Phi]+S_{\rm int}[\Phi]$, the partition function and n-point functions have  path integral representations (see \eg \cite{Ryder:1985wq}), \ie 
\beqa
Z\,=\,<0,+\infty|0,-\infty>&=&\int D\Phi e^{-S[\Phi]}\nn\\
<0,+\infty|\Phi_1(x_1)\cdots\Phi_n(x_n)|0,-\infty>&=&\int D\Phi\, (\Phi_1(x_1)\cdots \Phi_n(x_n))  e^{-S[\Phi]}\labell{ampf}
\eeqa
If the action has gauge symmetry, then one must use the Faddeev-Popov gauge fixing mechanism      to find finite result for the partition function and for the n-point functions. In principle, the   path integral may be evaluated for any coupling constant. The result would be a function of the coupling constant which may then be power expanded to produce perturbative contributions in which the coupling constant appears with positive powers,  and non-perturbative contributions in which the coupling constant appears with negative powers. The perturbative contributions can be found by expanding $e^{-S_{\rm int}}$ to produce  the tree-level and m-loop-level Feynman diagrams and  then evaluating the  corresponding Feynman amplitudes using the free theory propagators. At weak coupling, the tree-level contribution is   larger than one-loop-level contribution, one-loop-level is   larger than two-loop-level, and so on. So the first few terms of the Feynman amplitudes are adequate for evaluating the n-point functions.  At strong couplings, however, (m+1)-loop-level contribution is larger than m-loop-level, so one has to consider the contribution of all loops to evaluate the n-point functions. The non-perturbative contributions, on the other hand,    have no  Feynman diagram representation. Sometimes some of these contributions can be found by finding saddle points of the path integral, as in the study of instantons.

In string theory, the perturbative contributions to the partition function have been formulated as path integral by Polyakov \cite{Polyakov:1981rd,Polyakov:1981re,Alvarez:1982zi,Becker:2007zj}, \eg in the bosonic oriented closed string theory it is given as
\beqa
Z\,=\,<0,+\infty|0,-\infty>&\sim&\int Dh_{\alpha\beta}DX^{\mu} e^{-S[h_{\alpha\beta},X^{\mu},\cdots]}
\eeqa
where dots in the world-sheet action represent background fields. The asymptotic value of the   dilaton represents the closed string coupling constant, \ie $g_s=e^{\phi_0}$.  In the partition function,  $S$ is the world-sheet action of free closed string. The path integral over the world-sheet metric $h_{\alpha\beta}$ means sum over all two-dimensional Riemann surfaces which are analog of the Feynman diagrams.  However, $S$ is invariant under world-sheet diffeomorphism and Weyl transformations which makes the partition function to be infinite. To find the finite physical result for the partition function, one should fix these symmetries by summing over Riemann surfaces which are not related to each other by diffeomorphism and Weyl transformations. In the conformal gauge, \ie $
h_{\alpha\beta}=e^{\psi}\eta_{\alpha\beta}$, 
the Faddeev-Popov gauge fixing mechanism produces the ghosts $b,c$. Dropping the volume of the diffeomorphism  group, the finite partition function then becomes
\beqa
Z&=&\int DbDc D\psi D X^{\mu} e^{-S[b,c,\psi,X^{\mu},\cdots]}
\eeqa
The path integral over the world-sheet fields $b,c$ now gives the sum over topologies of the Riemann surfaces, \ie each topology has a specific contribution to $b,c$. The path integral over the world-sheet field $\psi$, on the other hand,  integrates over all conformally inequivalent Riemann surfaces at each topology, \ie integrate over the moduli space ${\cal M}_{n_h}$ of the Riemann surface with genus $n_h$. The domain of this integral depends on topology of the Riemann surfaces, \eg for sphere $\psi=0$ because all spheres are conformally equivalent. The dimension of this integral is zero for sphere, is two for torus and is $2(3n_h-3)$ for Riemann surfaces with genus $n_h$. The partition function, then can be written as
\beqa
Z&=&\sum_{n_h=0}^{\infty}g_s^{2n_h-2}\int_{{\cal M}_{n_h}}Z_{n_h}\labell{Z3}
\eeqa
where the asymptotic value of the dilaton in $Z_{n_h}$ is zero, \ie the world-sheet action in the presence of constant   dilaton  $\phi_0$  which is $S_{\rm dil}=\phi_0 \chi({\cal M}_{n_h})=\phi_0(2-2n_h)$ has been extracted from $Z_{n_h}$. For the free theory whose world-sheet is a cylinder from $-\infty$ to $+\infty$ there is no coupling constant.

The S-matrix elements of N states in the bosonic oriented closed string theory are then given as
\beqa
A(1,2,\cdots, N)&\sim&g_s^N\int DbDc D\psi D X^{\mu} (V_1V_2\cdots V_n) e^{-S[b,c,\psi,X^{\mu},\cdots]}\nn\\
&=&\sum_{n_h=0}^{\infty}g_s^{2n_h-2+N}\int_{{\cal M}_{n_h }}<V_1V_2\cdots V_N>\labell{Amp}
\eeqa
where $V$'s are the conformal invariant vertex operators corresponding to the particle states, \eg the vertex operator corresponding to the ground state  $|0,p>$ with momentum $p^{\mu}$ is 
\beqa
 c(z)\bc(\bz)e^{ip\cdot X}&{\rm or}& \int d^2z e^{ip\cdot X}\labell{tach}
\eeqa
 The particle states and their corresponding vertex operators must satisfy the Virasoro constraints, \eg they give the on-shell relation $p^2=-m^2=2$ for the above vertex operator. We will note fix the numerical normalization of the vertex operators. They appear as an  overall numerical factor in the amplitude \reef{Amp}. One may fix the overall numerical factor of the scattering amplitude by comparing the amplitude at low energy with the corresponding amplitude constructed from the standard low energy effective actions.

Both forms of the vertex operators \reef{tach} may appear in the scattering amplitude. The dimension of the moduli space of the Riemann surfaces with genus $n_h$ and $N$ punctures ${\cal M}_{n_h,N }$ is $2(3n_h-3+N)$. So for Riemann surfaces with $n_h>1$, one has to use the integral form of the vertex operators because the dimension of ${\cal M}_{n_h }$ is $2(3n_h-3)$ and each vertex operator has a two dimensional integral. For sphere, the dimension of ${\cal M}_{n_h }$ is zero whereas the dimension of ${\cal M}_{n_h,N }$ is $2(N-3)$. So one has to use three vertex operators with ghost and the other operators in the integral form. Similarly, for the tours, the dimension of ${\cal M}_{n_h }$ is two whereas the dimension of ${\cal M}_{n_h,N }$ is $2(N)$, so one has to use one vertex operator with ghost and all others in the integral form. Alternatively, one may use only the integral form of the vertex operators, then the integrand of the amplitude should be invariant under a group with 6 parameters for the sphere, \ie $SL(2,C)$, and   a group with two parameters for the tours. These symmetries should be fixed by fixing the position of three vertex operators in sphere and the position of one vertex operator in the tours. After taking into account the proper Jacobian factor which    is in fact the contribution of the ghost if one would use the vertex operator with ghost, the volume of these groups should be removed from the amplitude.     We will use this latter approach for calculation the scattering amplitudes.

If one includes  non-perturbative D$_p$-brane or  O$_p$-plane objects with $p<25$ in the bosonic string theory, then   the scattering amplitude of N closed string with $n_b$ D$_p$-branes and $n_c$ O$_p$-planes is given by \reef{Amp} in which the two-dimensional surfaces  have $n_b$ boundaries and  $n_c$ cross-caps, \ie
\beqa
A(1,2,\cdots, N,n_b,n_c) &\sim&\sum_{n_h=0}^{\infty}g_s^{2n_h+n_b+n_c-2+N}\int_{{\cal M}_{n_h,n_b,n_c }}<V_1V_2\cdots V_N>\labell{AmpDO}
\eeqa
For the D$_p$-brane, one should also consider open string vertex operators   at the boundaries of the two-dimensional surfaces which represent perturbative excitations of the D$_p$-brane. The dimension of D$_p$-brane is specified by imposing Newman or Dirichlet boundary condition on the world-sheet fields.     Unitarity requires the open string coupling constant to be related to the closed string coupling constant as  $g_o^2=g_s$. The above amplitude represents also the scattering amplitude in   superstring theories. For the superstrings, however, one should use the appropriate vertex operators.  
Using the Wick theorem, one can   calculate the correlation functions in   \reef{AmpDO}  by using the appropriate world-sheet propagators. 

One may also consider the scattering amplitude of closed string vertex operators on the world-volume of D$_p$-brane in the presence of constant background B-field. This can be   included into the amplitude \reef{AmpDO} by imposing mixed boundary conditions on the world-volume directions along which the B-field is non-zero \cite{Arfaei:1997hb,Garousi:1998bj}. For the open string states, however, the world-volume of D$_p$-brane in the presence of constant B-field remains ordinary commutative space in the Pauli-Villars regularization which is used in the non-linear sigma-model approach to the effective action, whereas it becomes non-commutative space in the point-splitting regularization which is used in the S-matrix approach to the effective action \cite{Seiberg:1999vs}, \ie the open string vertex operators in the presence of B-field correspond to the non-commutative fields.  As a result, there are two different   open string gauge fields. One corresponds to ordinary gauge symmetry and the other one correspond to non-commutative gauge symmetry. The differential equation which maps   these two variables, has been found by Seiberg and Witten by requiring the ordinary DBI action in the presence of constant B-field to be mapped to  non-commutative DBI action \cite{Seiberg:1999vs}.  

\subsection{Sphere-level amplitude of four tachyons in bosonic theory}

In this section we are going to calculate   the tree-level scattering amplitude of four tachyons in the bosonic string theory to show how one can explicitly derive the  Virasoro-Shapiro amplitude \cite{Virasoro:1969me,Shapiro:1969km}. The scattering amplitude \reef{AmpDO} for four closed string tachyon vertex operators \reef{tach} at sphere level is
\beqa
A(1,2,3,4)&\sim&g_s^2<V_1V_2V_3V_4>\labell{A1234}
\eeqa
To perform the correlators, one needs the   propagators of the world-sheet fields $X^{\mu}(\tau,\sigma)$ on the sphere. Since the amplitude is invariant under conformal transformation, one may perform conformal transformation to map the sphere to the complex plane. The   propagator of  $X^{\mu} $ on the complex plane is\footnote{Our conventions set $\alpha'=2$.}
\beqa
<X^{\mu}(z)X^{\nu}(w)>&=&-\eta^{\mu\nu}\log(z-w)\nn\\
<\bX^{\mu}(\bz)\bX^{\nu}(\bw)>&=&-\eta^{\mu\nu}\log(\bz-\bw)\nn\\
<\bX^{\mu}(\bz)X^{\nu}(w)>&=&0\labell{pro1}
\eeqa
where $X^{\mu}(z )+\bX^{\mu}(\bz)=X^{\mu}(z,\bz)$. Since there is no propagator between the holomorphic and the antiholomorphic part of $X^{\mu}$, the amplitude separates into holomorphic and antiholomorphic parts, \ie 
\beqa
A &\sim&g_s^2\int d^2z_1d^2z_2d^2z_3d^2z_4<\prod_{i=1}^4e^{ip_i\cdot X(z_i)}><\prod_{i=1}^4e^{ip_i\cdot \bX(\bz_i)}>
\eeqa
Using the following identity between the exponential of arbitrary operators $a(z)$, $b(w)$:
\beqa
:e^{a(z)}::e^{b(w)}:&=&:e^{a(z)+b(w)}:e^{<a(z)b(w)>}
\eeqa
where $:O:$ means normal order of the operator $O$,  one finds
\beqa
A &\sim&g_s^2\int d^2z_1d^2z_2d^2z_3d^2z_4\prod_{i<j}|z_i-z_j|^{2p_i\cdot p_j}\delta^{26}(p_1+p_2+p_3+p_4)
\eeqa
As we have already pointed out, the integrand must be invariant under the 6-parameter group $SL(2,C)$ because we have used the integral form of the vertex operators in \reef{A1234}. The $SL(2,C)$ transformation is
\beqa
z\rightarrow z'=\frac{az+b}{cz+d}&;&\bz\rightarrow \bz'=\frac{a^*\bz+b^*}{c^*\bz+d^*} 
\eeqa
where the complex parameters $a,b,c,d$ satisfy $ad-bc=1$. Under this transformation, one finds
\beqa
d^2z_i'=\frac{d^2z_i}{|cz_i+d|^4}&\& & |z_i'-z_j'|^2=\frac{|z_i-z_j|^2}{|cz_i+d|^2|cz_j+d|^2}
\eeqa
Using the on-shell condition $p_i^2=2$, one observes that the integrand is invariant under the $SL(2,C)$ transformation, so the amplitude becomes infinite. To avoid this infinity, one must fix  the $SL(2,C)$ symmetry. The infinitesimal form of the $SL(2,C)$ transformation is 
\beqa
\delta z&=&\alpha_1 +\alpha_2 z+\alpha_3 z^2\labell{delz}
\eeqa
where $\alpha_1,\alpha_2,\alpha_3$ are the three complex parameters of the group. One should use these three parameters to fix the position of three vertex operators at arbitrary points, \ie
\beqa
\int d^2z_1d^2z_2d^2z_3&=& \det \big[\frac{\prt(z_1,z_2,z_3)}{\prt(\alpha_1,\alpha_2,\alpha_3)}\big]\int d^2\alpha_1d^2\alpha_2d^2\alpha_3\,
\eeqa
Using \reef{delz}, one finds the Jacobian factor to be  $|(z_1-z_2)(z_2-z_3)(z_1-z_3)|^2$. The integral over the  parameters gives the volume of $SL(2,C)$ group which should be removed from the amplitude. Choosing $z_1=0,\, z_2=1,\, z_3=\infty,\, z_4\equiv z$, the amplitude then becomes
\beqa
A &\sim&g_2^2\int d^2|z|^{2p_1\cdot p_4}|1-z|^{2p_2\cdot p_4}\labell{amp1}
\eeqa
where we have also omitted for simplicity  the Dirac delta-function on the momenta. 
Using definition of the gamma function, \ie $\Gamma(\alpha)=\int_0^\infty dx x^{\alpha-1}e^{-x}$, one can write
\beqa
|z|^{-2\alpha}&=&\frac{1}{\Gamma(\alpha)}\int_0^\infty dx \,x^{\alpha-1}e^{-x|z|^2}\nn\\
|1-z|^{-2\beta}&=&\frac{1}{\Gamma(\beta)}\int_0^\infty dx \,x^{\beta-1}e^{-x|1-z|^2}
\eeqa
This turns the z-integration in \reef{amp1} into a Gaussian that can be explicitly carried out, \ie
\beqa
A &\sim&\frac{g_s^2}{\Gamma(-p_1\cdot p_4)\Ga(-p_2\cdot p_4)}\int_0^\infty dxdy\, x^{-p_1\cdot p_4-1}y^{-p_2\cdot p_4-1}\int d^2ze^{-x|z|^2-y|1-z|^2}
\eeqa
The Gaussian integral becomes
\beqa
\int d^2ze^{-x|z|^2-y|1-z|^2}&=&\frac{\pi}{x+y}e^{-\frac{xy}{x+y}}
\eeqa
Using the change of variables as $x=n/m,\, y=n/(1-m)$, one finds
\beqa
A&\sim&\frac{\pi g_s^2}{\Gamma(-p_1\cdot p_4)\Ga(-p_2\cdot p_4)}\int_0^1 dm(1-m)^{p_2\cdot p_4} m^{p_1\cdot p_4}\int_0^\infty dn n^{-(p_2\cdot p_4+p_1\cdot p_4+2)}e^{-n}
\eeqa
The last integral can be written in terms of the gamma function. Using the definition of the beta function, $B(\alpha,\beta)=\int_0^1 dx (1-x)^{\alpha-1}x^{\beta-1}$, one finds the following final result:
\beqa
A&\sim &\pi g_s^2\frac{\Ga(-t/2-1)\Ga(-s/2-1)\Ga(-u/2-1)}{\Ga(u/2+2)\Ga(s/2+2)\Ga(t/2+2)}
\eeqa
where we have also written the result in terms of the Mandelstam variables $t=-(p_1+p_2)^2$, $s=-(p_1+p_4)^2$ and $u=-(p_1+p_3)^2$. This is the Virasoro-Shapiro amplitude \cite{Virasoro:1969me,Shapiro:1969km}.

This amplitude has manifest symmetry amongst $t$-, $s$- and $u$-channels. From the poles of the gamma functions, one finds that the amplitude has simple poles at $s,t,u=-2,0,2,4,\cdots$. Since the poles of the tree-level scattering amplitudes correspond to the propagation of on-shell intermediate particles, one realizes that the mass of the intermediate particles are $m^2=-2,0,2,4,\cdots$. The first one corresponds to the closed string tachyon, the second one corresponds to the graviton and all others correspond to infinite tower of massive closed string states. 

\subsection{Sphere-level amplitude of four gravitons in type II theory}

In this section, we are going to review  the calculation of  the scattering amplitude of four gravitons at sphere-level and discuss how higher derivative couplings of gravitons in type II superstring theory can be found from the scattering amplitude.

In  superstring theory,  the world-sheets carry background charge of the superghost field $\phi$ (see \eg \cite{Peskin:1987rz}).  The sphere in type II theory carries independent  background charge for  holomorphic and for antiholomorphi part of $\phi$, \ie $Q_{\phi}=(2,2)$. These   charges must be compensated by the vertex operators in the scattering amplitude. In fact, in superstring theory,   a given physical state can carry different superghost charges. As a result, there are different vertex operators corresponding to a  given physical state.   One must choose the vertex operators in a scattering amplitude such that they compensate the background charge of $\phi$. The scattering amplitude \reef{AmpDO} for four  superstring graviton vertex operators  at sphere-level is
\beqa
A(1,2,3,4)&\sim&g_s^2<V^{(0,0)}_1V^{(0,0)}_2V^{(-1,-1)}_3V^{(-1,-1)}_4>\labell{amp2}
\eeqa
where the superscripts represent the superghost charges.  The graviton vertex operator in pictures $(0,0)$ and $(-1,-1)$ are given as (see \eg \cite{Peskin:1987rz})
\beqa
V^{(0,0)}&=&\veps_{\mu\nu}\int d^2z:(\prt X^{\mu}+ip\cdot  \psi\psi^{\mu})e^{ip\cdot X}:(\bprt \bX^{\nu}+ip\cdot\bpsi\bpsi^{\nu})e^{ip\cdot\bX}:\nn\\
V^{(-1,-1)}&=&\veps_{\mu\nu}\int d^2z: e^{-\phi}\psi^{\mu}e^{ip\cdot X}: e^{-\bphi}\bpsi^{\nu}e^{ip\cdot\bX}:\labell{ver}
\eeqa
The vertex operators satisfy the Virasoro constraint provided that the momentum and the polarization tensor satisfy the on-shell relations $p^2=0=p^{\mu}\veps_{\mu\nu}$. The propagators for the $X^{\mu}$ and $\phi$  are the same as \reef{pro1}, and the propagators for $\psi^{\mu}$ are
\beqa
<\psi^{\mu}(z)\psi^{\nu}(w)>&=&-\frac{\eta^{\mu\nu}}{z-w}\nn\\
<\bpsi^{\mu}(\bz)\bpsi^{\nu}(\bw)>&=&-\frac{\eta^{\mu\nu}}{\bz-\bw}\nn\\
<\psi^{\mu}(z)\bpsi^{\nu}(\bw)>&=&0\labell{pro11}
\eeqa
 Since there is no propagator between holomorphic and antiholomorphic, one may write $\veps_{\mu\nu}=\xi_{\mu}\bxi_{\nu}$ and separates the amplitude to holomorphic and antiholomorphic parts. Then using the above propagators, one can perform   different correlators  in \reef{amp2} and show that the integrand is invariant under $SL(2,R)\times SL(2,R)$ transformations. Fixing this symmetry, one can write the result in terms of the gamma functions as in the previous section. The result is \cite{Schwarz:1982jn}
\beqa
A&\sim&g_s^2K\bK\frac{\Ga(-t/2)\Ga(-s/2)\Ga(-u/2)}{\Ga(u/2+1)\Ga(s/2+1)\Ga(t/2+1)}\delta^{10}(\sum_{i=1}^4p_i)\labell{amp3}
\eeqa
where the kinematic factor $K$ is
\beqa
K&=&-p_2\cdot p_3p_2\cdot p_4\xi_1\cdot\xi_2\xi_3\cdot\xi_4-p_1\cdot p_2\big[\xi_1\cdot p_4\xi_3\cdot p_2 \xi_2\cdot\xi_4+\xi_2\cdot p_3\xi_4\cdot p_1 \xi_1\cdot \xi_3\labell{kin}\\
&+&\xi_1\cdot p_4\xi_4\cdot p_2 \xi_2\cdot\xi_3+\xi_2\cdot p_4\xi_3\cdot p_1 \xi_1\cdot\xi_4\big]+\{1,2,3,4\rightarrow 1,3,2,4\}+\{1,2,3,4\rightarrow 1,4,3,2\}\nn
\eeqa
Similarly for $\bK$. Using the identity $\Ga(x)\Ga(1-x)=\pi/\sin(\pi x)$,     the amplitude may be written in terms of the product of two disk-level scattering amplitude of four gauge bosons, one corresponds to the  holomorphic part and the other one corresponds to the antiholomorphic part, \ie
\beqa
A&\sim&\frac{\sin(-\pi t/2)}{\pi}\left(g_s K\frac{\Ga(-t/2)\Ga(-s/2)}{\Ga(1+u/2)}\right)\left(g_s \bK\frac{\Ga(-t/2)\Ga(-u/2)}{\Ga(1+s/2)}\right)\delta^{10}(\sum_{i=1}^4p_i)\nn
\eeqa
where the  first parenthesis is the disk-level scattering amplitude of four gauge bosons in $t,s$-channel, and the second one in $t,u$-channel \cite{Schwarz:1982jn}. Such relation which is known as  Kawai-Lewellen-Tye (KLT)  relation \cite{Kawai:1985xq}, is expected to be holed for all other closed string amplitudes.

In the amplitude \reef{amp3},   $\alpha'=2$. Using the fact that the amplitude should be dimensionless, one can restore the $\alpha'$ factors. Up to the  overall numerical factor, the amplitude can be written as  
\beqa
A
&\sim&\frac{g_s^4\alpha'^3}{2\kappa_{10}^2}K\bK\frac{\Ga(-t/2)\Ga(-s/2)\Ga(-u/2)}{\Ga(u/2+1)\Ga(s/2+1)\Ga(t/2+1)}\delta^{10}(\sum_{i=1}^4p_i)\labell{amp31}
\eeqa
where $2\kappa_{10}^2=g_s^2(2\pi\sqrt{\alpha'})^8/2\pi=2\kappa^2g_s^2$, and the Mandelstam variable $t$ become $t=-\alpha'(p_1+p_2)^2/2$. Similarly for all other Mandelstam variables. From the poles of the gamma functions in \reef{amp31}, one finds that the amplitude has simple poles at $s,t,u=0,2,4,\cdots$ which correspond to the  masses of the intermediate closed string states   $m^2=0,2,4,\cdots$. String field theory (see \eg \cite{Sen:2016qap}) which contains iteration of graviton with all of the massive states  should produce the amplitude \reef{amp3}  and any other  amplitude  \reef{AmpDO} which the perturbative  string  theory   produces . 

One may expand the gamma functions in \reef{amp31} around $\alpha'\rightarrow 0$, \ie
\beqa
A&\sim&\frac{g_s^4\alpha'^3}{2\kappa_{10}^2}K\bK\left(-\frac{8}{stu}-2\z(3)+\cdots\right)\delta^{10}(\sum_{i=1}^4p_i)\labell{amp4}
\eeqa
where dots represent terms with higher orders of the Mandelstam variables, \ie higher order of $\alpha'$.  The amplitude  has now simple poles at $s,t,u=0$ and    contact terms with infinite number of momenta. There should be an action which contains gravitons with infinite number of higher derivatives which produces the amplitude \reef{amp4} and produces $\alpha'$-expansion   of any other amplitude in perturbative string theory.

To find the relation between this action and the string field theory action, let us  denote the (finite number of )   massless fields collectively as $\Phi_0$, and the (infinite number of ) heavy fields collectively as $\Phi_H$. The appropriate string field theory should describe  the string theory by a classical action $S[\Phi_0,\Phi_H]$ governing these fields and their couplings. One may integrate out all massive fields to find a Wilsonian effective action $S_{\rm eff}[\Phi_0]$ in terms of only massless fields. The effect of massive fields appear as  infinite number of higher derivatives on the massless fields in the effective action, \ie
\beqa
e^{-S_{\rm eff}[\Phi_0]}=\int D\Phi_H e^{-S[\Phi_0, \Phi_H]}
\eeqa
 While the n-point function \reef{ampf} with the string field theory $S[\Phi_0,\Phi_H]$ should reproduce the amplitude \reef{amp3}, the n-point function \reef{ampf} with the Wilsonian effective action $S_{\rm eff}[\Phi_0]$ should reproduce the amplitude \reef{amp4}. 

The effective action has the following expansion:
\beqa
S_{\rm eff}[\Phi_0]&=&\sum_{n=0}^{\infty} \alpha'^nS_n[\Phi_0]\labell{eff}
\eeqa
The first term of the expansion, \ie $S_0[\Phi_0]$, is known for all string theories, \eg in type IIB it is given by \reef{sugraB}. Unlike the leading order action which has neither genus nor non-perturbative contributions, the actions at higher orders have both genus and non-perturbative corrections. If would be extremely hard to find the  effective action completely.   Even the action at the next to the leading order in which we are interested is not completely known.   

 The sphere-level scattering amplitude of any other four massless closed string vertex operators  in type II supergravity has the same structure as \reef{amp3}. The kinematic factor $K\bK$, however, depends on external states. So all such amplitudes have the same expansion as \reef{amp4}.   The first term in \reef{amp4} which is at the zeroth order of $\alpha'$, are   reproduced by $S_0[\Phi_0]$ \cite{Sannan:1986tz,Bakhtiarizadeh:2013zia,Bakhtiarizadeh:2015exa} and the second term which is at order $\alpha'^3$, should produce four-field couplings at eight-derivative level. 

The kinematic factor $K$ in \reef{kin} has four derivatives. If one considers $\xi$'s as the polarization tensors of open string  gauge bosons, then $K$ produces four gauge field strength couplings as $t_8F^4$ where the tensor $t_8$ is antisymmetric within a pair of indices and is symmetric under exchange of the pair of indices \cite{Schwarz:1982jn}, \ie for four arbitrary antisymmetric matrices $M^1,\,\cdots, M^4$ it is given as 
\beqa
&&\frac{1}{8}t_{8}M^1 M^2 M^3 M^4 =-\bigg[\Tr( M^1M^2M^3M^4)+\Tr (M^1M^3M^2M^4)+\Tr (M^1M^3M^4M^2)\bigg]\nonumber\\
&&\quad +\frac{1}{4}\bigg[\Tr (M^1M^2)\Tr (M^3M^4)+\Tr (M^1M^3)\Tr (M^2M^4)+\Tr (M^1M^4)\Tr (M^2M^3)\bigg]\labell{t8}
\eeqa
In terms of this tensor, the second term in \reef{amp4} produces   $t_8t_8R^4$ for   four-Riemann curvature couplings \cite{Gross:1986iv}. There is another four-Riemann curvature coupling, \ie $\eps_8\eps_8 R^4$ which is total derivative at four-metric level so its presence in the effective action could not be confirmed by four-gravitons amplitude.  The coefficient of this term has been found   from the non-linear $\sigma$-model   approach \cite{Grisaru:1986vi,Freeman:1986zh} to be $1/4$ with respect to the first term. It has been recently confirmed  this term is consistent with the sphere-level S-matrix element of five graviton vertex operators \cite{Garousi:2013tca}. 

The   B-field and dilaton couplings at four-field level have been added to  $t_8t_8R^4$    by extending the Riemann curvature in the Einstein frame to the generalized Riemann curvature at the linear order \cite{Gross:1986mw}\footnote{Note that the normalizations of the dilation  and B-field here  are $\sqrt{2}$ and 2 times the normalization of the dilaton and B-field in \cite{Gross:1986mw}, respectively.}, \ie
 \beqa
 \bar{R}_{\mu\nu}{}^{\alpha\beta}&= &R_{\mu\nu}{}^{\alpha\beta}- \eta_{[\mu}{}^{[\alpha}\phi_{,\nu]}{}^{\beta]}+  e^{-\phi/2}H_{\mu\nu}{}^{[\alpha,\beta]}\labell{trans1}
\eeqa
where   the bracket notation is $H_{\mu\nu}{}^{[\alpha,\beta]}=\frac{1}{2}(H_{\mu\nu}{}^{\alpha,\beta}-H_{\mu\nu}{}^{\beta,\alpha})$, and comma denotes the partial derivative. 
Using the relation between the Einstein  frame metric and the string frame metric $G_{\mu\nu}=e^{-\phi/2}G^s_{\mu\nu}$, one observes that the dilaton term in above equation is canceled in   transforming the linearized  Riemann curvature from the Einstein frame to the string frame   \cite{Garousi:2012jp}, \ie
\beqa
\bar{R}_{\mu\nu \alpha\beta}&\Longrightarrow&e^{-\phi/2}\cR_{\mu\nu \alpha\beta}
\eeqa
where on the right hand side the metric is in the string frame. In above equation,  $\cR_{\mu\nu \alpha\beta}$ is the following expression 
\beqa
\cR_{\mu\nu \alpha\beta }&=&R_{\mu\nu \alpha\beta}+H_{\mu\nu [\alpha;\beta]}\labell{RH2}
\eeqa
 where we have also extended the ordinary derivative to the covariant derivative, and the linearized Riemann curvature to the covariant Riemann curvature.  The  action involving   Riemann curvature, $\nabla H$  and $\nabla\nabla\phi$   at the sphere level then becomes
\beqa
S_{II}\supset \frac{\gamma \z(3)}{3.2^7\kappa^2} \int d^{10}x e^{-2\phi} \sqrt{-G}(t_8t_8\cR^4+\frac{1}{4}\eps_{8}\eps_{8} R^4)\labell{Y3}
\eeqa
where $\gamma=\frac{\alpha'^3}{2^{5}}$ and the metric is in the string frame.  The above action, however, does not include couplings involving $H$ and $\nabla \phi$. It does not include RR fields either. In principle, all these couplings may be found by extracting the corresponding sphere-level S-matrix elements. For example, the couplings involving only $H$ at eight-derivative level have structure $H^8$. These couplings may be found by analyzing the sphere-level S-matrix element of eight vertex operators. However, it is extremely difficult to calculate such S-matrix element and to extract its eight-momentum contact terms. Even worse, the action has   genus and non-perturbative corrections. The latter can not be found from the perturbative S-matrix elements \reef{AmpDO}. So we have to use another technique to find such couplings. One expects the consistency of the couplings found in \reef{Y3} with dualities may fix all other couplings. 

 \subsection{Disk-level amplitude of two closed strings and its $\alpha'^2$-couplings}

In this section, we are going to review  the calculation of  the scattering amplitude of two massless closed strings  at disk-level in type II superstring theory and discuss how higher derivative couplings   in the DBI and WZ actions can be found from the scattering amplitudes.

  In type II theory    the  background charge of disk is $Q_{\phi}= 2$.   The S-matrix element \reef{AmpDO} for two    gravitons    at disk-level which represents the  scattering amplitude of one graviton off a D$_p$-brane, is
\beqa
A(1,2 )&\sim&g_s<V^{(0,0)}_1 V^{(-1,-1)}_2 >\labell{ampd}
\eeqa
where the graviton vertex operators are given in \reef{ver}. The vertex operators for Kalb-Ramond B-field and for dilaton are the same as  \reef{ver} in which the polarization tensor is antisymmetric for B-field and is $\veps_{\mu\nu}\sim \eta_{\mu\nu}+p_{\mu}\ell_{\nu}+p_{\nu}\ell_{\mu}$ where the auxillary vector $\ell_{\mu}$ satisfies $\ell\cdot p=1$, for dilation. The dimension of D$_p$-brane is specified by imposing Newman or Dirichlet boundary condition on world sheet fields $X^{\mu},\psi^{\mu}$. Because of the boundary conditions, the   holomorphic and   antiholomorphi part of fields on disk are not independent. The propagators between holomorphic fields and between antiholomorphic fields are the same as the sphere propagators, however, the propagators between holomorphic and antiholomorphic fields depend on boundary conditions on fields. Using conformal transformation to map disk to the upper-half plane, one finds the standard propagators \reef{pro1} and \reef{pro11} between holomorphic fields and between antiholomorphic fields, and the following propagators between holomorphic and antiholomorphic fields:
\beqa
<X^{\mu}(z)\bX^{\nu}(\bw)>&=&-D^{\mu\nu}\log(z-\bw)\nn\\
<\psi^{\mu}(z)\bpsi^{\nu}(\bw)>&=&-\frac{D^{\mu\nu}}{z-\bw}\nn\\
<\phi(z)\bphi(\bw)>&=&- \log(z-\bw)\labell{pro2}
\eeqa
where matrix $D^{\mu\nu}=\eta^{\mu\nu}$ for the directions that the Newman boundary condition is imposed and  $D^{\mu\nu}=-\eta^{\mu\nu}$ for the directions that the Dirichlet  boundary condition is imposed, \ie for D$_p$-brane it is $D_{\mu}{}^{\nu}={\rm diag}(\underbrace{1,1,\cdots, 1}_{p+1}, -1,-1,\cdots,-1)$. One may use the doubling trick \cite{Garousi:1996ad} 
\beqa
\bX^{\mu}(\bz)\rightarrow D^{\mu}{}_{\nu} X^{\nu}(\bz)\,,\,\,\,\, \bpsi^{\mu}(\bz)\rightarrow D^{\mu}{}_{\nu} \psi^{\nu}(\bz)\,,\,\,\,\, \bphi (\bz)\rightarrow \phi(\bz)\labell{trick}
\eeqa
to remove the matrix $D^{\mu\nu}$ from the propagators \reef{pro2}. The above replacement makes the antiholomorphic fields  in the vertex operator  \reef{ver} to be in terms of holomorphic fields, the momentum in the antiholomorphic part to be $p\inn D$ and the overall polarization tensor to be $(\veps\inn D)_{\mu\nu}$.

The dimension of the moduli space of disk with two punctures is one, so the integrand in \reef{ampd} must be invariant under a group with three real parameter. In fact, using the above propagators, one can perform   different correlators  in \reef{ampd} and show that the integrand is invariant under $SL(2,R)$ transformations. Fixing this symmetry, by setting $z_1=i$ and $z_2-iy$, \ie
\beqa
\int  d^2z_1d^2z_2&\rightarrow & \int_0^1 dy (1-y^2)
\eeqa
one can write the result in terms of the gamma functions as in the previous sections. The result is \cite{Garousi:1996ad} 
\beqa
A&\sim&\alpha'^2g_s^2T_p\, K(1,2)\frac{\Ga(-t/2)\Ga(-2q^2)}{\Ga(1-t/2-2q^2)}\delta^{p+1}(p_1\inn V+p_2\inn V)\labell{ampd2}
\eeqa
where
$t=-\alpha'(p_1+p_2)^2/2$ is the momentum transfer to the
 D$_p$-brane, and $q^2=-\alpha'(p_1\inn V)^2/2$ is the momentum flowing parallel to the world-volume of
the  D$_p$-brane. In above amplitude,  we have also restored the $\alpha'$ dependence by using the fact that the amplitude should be dimensionless. In the amplitude $T_p=\frac{1}{g_s(2\pi)^p(\sqrt{\alpha'})^{p+1}}$ is the D$_p$-brane tension in the string frame and the matrix $V^{\mu\nu}$ is  $V_{\mu}{}^{\nu}={\rm diag}(\underbrace{1,1,\cdots, 1}_{p+1}, 0,0,\cdots,0)$. Note that for the convention $\alpha'=2$, $\alpha'^2g_s^2T_p=g_s$, up to a numerical factor. The kinematic factor $K(1,2)$ is 
\beqa
K(1,2)&=&2q^2 a_1+\frac{t}{2}a_2
\eeqa
and the explicit form of $a_1,\, a_2$ is \cite{Garousi:1996ad} 
\beqa
 a_1&=& {\rm
Tr}(\pol_1\inn D)\,p_1\inn \pol_2 \inn p_1 -p_1\inn\pol_2\inn
D\inn\pol_1\inn p_2 - p_1\inn\pol_2\inn\pol_1^T \inn D\inn
p_1  -p_1\inn\pol_2^T \inn \pol_1 \inn D \inn p_1
\nonumber\\
&&  -
\frac{1}{2}(p_2\inn\pol_1^T\inn\pol_2 \inn p_1+
p_1\inn\pol_2\inn\pol_1^T \inn p_2)  +\frac{1}{2}p_1\inn D\inn p_1 \Tr(\pol_1\inn  \pol_2^T)
+\Big\{1\longleftrightarrow
2\Big\} \nonumber\\
a_2&=&{\rm Tr}(\pol_1\inn D)\,(p_1\inn\pol_2\inn D\inn p_2-
p_2\inn D\inn\pol_2\inn D\inn p_1 )
\nonumber\\
&&+p_1\inn D\inn\pol_1\inn D\inn\pol_2\inn D\inn p_2 -p_2\inn
D\inn\pol_2\inn\pol_1^T\inn D\inn p_1+\frac{1}{2}p_1\inn D\inn p_1\Tr(\pol_1\inn D\inn\pol_2\inn D)
\nonumber\\
&& -\frac{1}{2}p_1\inn D\inn p_1 \Tr(\pol_1\inn  \pol_2^T)-\frac{1}{2}{\rm Tr}(\pol_1\inn D) {\rm Tr}(\pol_2\inn D)\,(p_1\inn D\inn p_1+p_1\inn p_2)
+\Big\{1\longleftrightarrow 2 \Big\} \labell{fintwo} 
\eeqa 
  The disk-level scattering amplitude of any other two massless closed string vertex operators  in type II supergravity has the same structure as \reef{ampd2}. The kinematic factor $K(1,2)$, however, depends on external states. We refer the interested readers to \cite{ Garousi:1996ad}  for the explicit form of the kinematic factor for all other states. A constant background B-field can be added to the scattering amplitude \reef{ampd2} by extending the diagonal matrices $D^{\mu\nu}$ and $V^{\mu\nu}$ to   non-diagonal matrices which include the background B-field \cite{Garousi:1998bj}. 

 From the poles of the gamma functions in \reef{ampd2}, one finds that the amplitude has simple poles at $t=0,2,4,\cdots$ in the $t$-channel which correspond to the  masses of the intermediate closed string states   $m^2=0,2,4,\cdots$, and at $q^2=0,1/2,1,3/2,\cdots$ in the $q^2$-channel which correspond to the masses of the intermediate open string states $m_{\rm open}^2=0,1/2,1,3/2,\cdots$.  The world-volume string field theory   which contains iteration of graviton with all of the massive open and closed states  should produce the amplitude \reef{ampd2}.

One may expand the gamma functions in \reef{ampd2} around $\alpha'\rightarrow 0$, \ie
\beqa
A&\sim&\alpha'^2g_s^2T_p\, K(1,2)\left(-\frac{1}{tq^2}-\frac{\pi^2}{6}+\cdots\right)\delta^{p+1}(p_1\inn V+p_2\inn V)\labell{ampd4}
\eeqa
where dots represent terms with higher orders   $\alpha'$.  The amplitude  has now simple poles at $q^2,t=0$ and    contact terms with infinite number of momenta. The simple poles are reproduced by the DBI action   and the supergravity. The higher derivative extension  of the DBI action should  produce the contact terms of the amplitude \reef{ampd4}.

The amplitude \reef{ampd4} at order $\alpha'^2$ has only contact term, \ie
\beqa
A(\alpha'^2)&\sim&-\frac{\pi^2}{6}\alpha'^2g_s^2T_p\, K(1,2) \delta^{p+1}(p_1\inn V+p_2\inn V)\labell{ampd44}
\eeqa
It must be reproduced by some covariant couplings at order $\alpha'^2$ which include the two massless NS-NS fields. That is, one has to write all such couplings with unknown coefficients, and then transform their corresponding two-field couplings to the momentum space and use the on-shell conditions. The result should be the same as \reef{ampd44}. This constrains   the unknown coefficients. 

In this way, the  higher derivative corrections at order  $\alpha'^2$ involving Riemann curvature, the second fundamental form, $\nabla H$ and $\nabla\nabla \phi$ have  been found      
in \cite{Bachas:1999um,Garousi:2009dj,Garousi:2011fc,Garousi:2009dj} to be 
\beqa 
S_p^{DBI} &\supset&-\frac{\pi^2\alpha'^2T_{p}}{48}\int d^{p+1}x\,e^{-\phi}\sqrt{-\tG}\bigg[(R_T)_{abcd}(R_T)^{abcd}-2\bar{\cR}_{ab}\bar{\cR}^{ab}-(R_N)_{abij}(R_N)^{abij}\nn\\
&&+2\bar{\cR}_{ij}\bar{\cR}^{ij}+\frac{1}{2}\nabla_aH_{bci}\nabla^aH^{bci}-\frac{1}{6}\nabla_aH_{ijk}\nabla^aH^{ijk}-\frac{1}{3}\nabla_iH_{abc}\nabla^iH^{abc}
 \bigg]\labell{RTN}
\eeqa
where $\tG$ is determinant of the pull-back metric, $\tG_{ab}=\prt_a X^{\mu}\prt_b X^{\nu}G_{\mu\nu}$,  and  
 the  curvatures $(R_T)_{abcd}$ and $(R_N)^{abij}$  are related to the projections of  the bulk Riemann curvatures into world-volume and transverse spaces, and to the   second fundamental form via the Gauss-Codazzi equations, \ie
\beqa
(R_T)_{abcd}&=&R_{abcd}+ \delta_{ij}(\Omega_{\ ac}{}^i\Omega_{\ bd}{}^j-\Omega_{\ ad}{}^i\Omega_{\ bc}{}^j)\nonumber\\
(R_N)_{ab}{}^{ ij}&=&R_{ab}{}^{ij}+\tG^{cd}(\Omega_{\ ac}{}^i\Omega_{\ bd}{}^j-\Omega_{\ ac}{}^j\Omega_{\ bd}{}^i)\labell{RTRN}
\eeqa
The   curvatures $\bar{\cR}_{ab}$ and $\bar{\cR}_{ij}$  are related to the   Riemann curvatures, the   second fundamental form and to the second derivative of dilaton via the following relations:
\beqa
\bar{\cR}_{ab}&=& R^c{}_{acb}+ \delta_{ij}(\Omega_c{}^c{}^i\Omega_{\ ab}{}^j-\Omega_{\ ca}{}^i\Omega_b{}^c{}^j)+\nabla_a\nabla_b\phi\nonumber\\
\bar{\cR}_{ij}&=& R^c{}_{icj}+ \delta_{ik}\delta_{jl} \Omega^{ab}{}^k\Omega_{ab}{}^l+\nabla_i\nabla_j\phi\labell{Rij}
\eeqa
The world-volume indices in \reef{Rij} and \reef{RTN} are raised by the inverse of the pull-back metric, and the transverse indices in \reef{RTN} are raised by $\delta^{ij}$. Note that if $A_{\mu}, B_{\mu}$ are two spacetime vectors, one can write  $A_aB^a=A_aB_b\tG^{ab}=A_{\mu}B_{\nu}\prt_a X^{\mu}\prt_b X^{\nu}\tG^{ab}=A_{\mu}B_{\nu}\tG^{\mu\nu}$ where the projection operator $\tG^{\mu\nu}$ is the first fundamental form, and $A_iB^i=A_iB_j\delta^{ij}=A_{\mu}B_{\nu}\z^{\mu}_i\z^{\nu}_j\delta^{ij}=A_{\mu}B_{\nu}\bot^{\mu\nu}$ where $\z^{\mu}_i$ is a  orthonormal frame for the normal bundle and $\bot^{\mu\nu}$ is a projection operator that projects spacetime tensors to the normal space. The two projections satisfy $\tG^{\mu\nu}+\bot^{\mu\nu}=G^{\mu\nu}$.  So one can write world-volume couplings either in terms of world-volume and transvers indices, or in terms of spacetimes indices in which the metric $G^{\mu\nu}$ and the first fundamental form are used to contract the indices. 

The second fundamental form is defined as the   covariant derivative of the tangent vectors $\prt_a X^{\mu}$, \ie $\Omega_{ab}{}^{\mu}=\nabla_a\prt_b X^{\mu}$ (see \eg the appendix in \cite{Bachas:1999um}). Using the relation $\nabla_a(A_bB^{\mu})=(\nabla_aA_b)B^{\mu}+A_b\prt_a X^{\nu}\nabla_{\nu}B^{\mu}$, one finds
\beqa
\Omega_{ab}{}^{\mu} 
&=&\prt_a\prt_b X^{\mu}-\tilde{\Ga}_{ab}{}^c\prt_c X^{\mu}+\Ga_{ab}{}^{\mu} 
\eeqa
where $\tilde{\Ga}_{ab}{}^c$ is the world-volume connection constructed from the pull-back metric and $\Ga_{ab}{}^{\mu} $ is  pull-back of the spacetime connection. Using the fact that $\nabla_a\tG_{bc}=0$ and $\nabla_aG_{\mu\nu}=\prt_a X^{\rho}\nabla_{\rho}G_{\mu\nu}=0$, one observes that the projection of the second fundamental form to the world-volume is zero. The projection of this tensor to the normal space which appears in \reef{RTRN} and \reef{Rij} is 
\beqa
\Omega_{ab}{}^i\equiv\Omega_{ab}{}^{\mu}\z^{\nu}_jG_{\mu\nu}\delta^{ij}
\eeqa
Note that the vectors $\z^{\mu}_i$ do not appear in the action, \ie $\delta_{ij}\Omega_{ab}{}^i\Omega_{cb}{}^j=\bot_{\mu\nu}\Omega_{ab}{}^{\mu}\Omega_{cd}{}^{\nu}$ where $\bot_{\mu\nu}=G_{\alpha\mu}G_{\beta\nu}\bot^{\alpha\beta}$. To relate the second fundamental form to the open string transvser scalars, one has to write the above covariant coupling in the static gauge in which $X^{a}=\sigma^a$ and the other components are the transverse scalar fields, \ie $X^i=\chi^i$. In this gauge, $\nabla_a\prt_b X^c=0$ and 
\beqa
\bot_{\mu\nu}\Omega_{ab}{}^{\mu}\Omega_{cd}{}^{\nu}&=&\bot_{ij}K_{ab}{}^iK_{cd}{}^j
\eeqa
where $K_{ab}{}^i$ is
\beqa
K_{ab}{}^i&=&\prt_a\prt_b\chi^i-\tilde{\Ga}_{ab}{}^c\prt_c\chi^i+\Ga_{ab}{}^i+\Ga_{aj}{}^i\prt_b\chi^j+\Ga_{bj}{}^i\prt_a\chi^j+\Ga_{jk}{}^i\prt_a\chi^j\prt_b\chi^k\labell{sec}
\eeqa
and $\Ga_{ab}{}^i$, $\Ga_{aj}{}^i$, $\Ga_{jk}{}^i$ are different components of the spacetime connection. In  finding the couplings of one massless closed and two open strings \cite{Bachas:1999um}, one considers only the first term in $K_{ab}{}^i$. All other terms should be reproduced by the contact terms of the higher S-matrix elements.

Analyzing the $\alpha'^2$-order terms in amplitude \reef{ampd44} for one RR and one NSNS vertex operators, the  higher derivative corrections to the WZ action   at order  $\alpha'^2$ involving one RR and one NSNS fields      have been found in   
\cite{Garousi:2010ki} to be
\beqa
S_p^{WS} &\!\!\!\!\!  \supset \!\!\!\!\!&-\frac{\pi^2\alpha'^2T_{p}}{24}\int d^{p+1}x\,\eps^{a_0\cdots a_p}\left(\frac{1}{3!(p+1)!}\nabla_a\cF^{(p+4)}_{ia_0\cdots a_pjk}\nabla^aH^{ijk}\right.\labell{LTdual}\\
&&\left.\qquad\qquad +\frac{2}{p!}[\frac{1}{2!}\nabla_a\cF^{(p+2)}_{ija_1\cdots a_p}(R_N)_{a_0}{}^{aij}+\frac{1}{p+1}\nabla_j\cF^{(p+2)}_{ia_0\cdots a_p}\bar{\cR}^{ij}]\right.\nonumber\\
&&\left.\qquad\qquad+\frac{1}{2!(p-1)!}[\nabla^a\cF^{(p)}_{ia_2\cdots a_p}\nabla^iH_{aa_0a_1}-\frac{1}{p}\nabla^i\cF^{(p)}_{a_1a_2\cdots a_p}(\nabla^aH_{iaa_0}-\nabla^jH_{ija_0})]\right)\nonumber
\eeqa
where $\cF^{(p)}$ is the linearized RR field strength, \ie  $\cF^{(p)}=dC^{p-1}$. The above couplings include $\cF^{(1)},\cdots, \cF^{(9)}$ where $\cF^{(9)}=\star_{10} \cF^{(1)}$, $\cF^{(8)}=\star_{10} \cF^{(2)}$, $\cF^{(7)}=\star_{10} \cF^{(3)}$, $\cF^{(6)}=\star_{10} \cF^{(4)}$ and $\cF^{(5)}=\star_{10} \cF^{(5)}$. Here also the world-volume indices   are raised by the inverse of the pull-back metric,  the transverse indices  are raised by $\delta^{ij}$, and the tensors with the lower indices are the projections of  the bulk tensors into world-volume and transverse spaces.

Analyzing the $\alpha'^2$-order terms in amplitude \reef{ampd44} for two RR   vertex operators, the  higher derivative corrections to the D$_p$-brane action   at order  $\alpha'^2$ involving two RR   fields      have been found in   
\cite{Garousi:2011fc} to be\footnote{Note that, there is a typo in $a_1$ in   equation above (12) in \cite{Garousi:2011fc} as an extra  $\star_{10}$ operator. Since in $a_1$ there is one projection operator $P_-$, there should be only one $\star_{10}$ in $a_1$.}
\beqa
S^{DBI}_{p}&\!\!\!\!\supset\!\!\!\!&-\frac{\pi^2\alpha'^2T_p}{96 }\int d^{p+1}x\,e^{\phi}\sqrt{-\tG}\left(\sum_{n=1}^4\frac{1}{n!}\bigg[(p-4)\nabla_a\cF^{(n)} \inn \nabla^a\cF^{(n) }-nD^{\mu}{}_{\nu}\nabla_a\cF^{(n)}{}_{\mu} \inn \nabla^a\cF^{(n)\nu} \right.\nonumber\\
&&+4\delta_{n,p}\nabla_{\mu}\cF^{(n)} \inn V\inn \nabla^{\mu}\cF^{(n) } +4n\delta_{n,p+2}\nabla_{\mu}\cF^{(n)}{}_{i }\inn V\inn \nabla^{\mu}\cF^{(n)\,i }\bigg]+\sum_{n=1}^4\frac{1}{(10-n)!}\bigg[\nonumber\\
&&4\delta_{10-n,p}\nabla_{\mu}\cF^{(10-n)} \inn V\inn \nabla^{\mu}\cF^{(10-n) } +4(10-n)\delta_{10-n,p+2}\nabla_{\mu}\cF^{(10-n)}{}_{i }\inn V\inn \nabla^{\mu}\cF^{(10-n)\,i }\bigg]\nn\\
&&+\frac{1}{ 5!}\bigg[(p-4)\nabla_a\cF^{(5)} \inn \nabla^a\tF^{(5) } -5D^{\mu}{}_{\nu}\nabla_a\cF^{(5)}{}_{\mu} \inn \nabla^a\tF^{(5)\nu}  \nonumber\\
&&\left.+4\delta_{5,p}\nabla_{\mu}\tF^{(5)}  \inn V\inn \nabla^{\mu}\tF^{(5) }  +20\delta_{5,p+2}\nabla_{\mu}\tF^{(5)}{}_i  \inn V\inn \nabla^{\mu}\tF^{(5)}{}^{i } \bigg]\right)\labell{FF}
\eeqa
where $\cF^{(10-n)}=\star_{10}\cF^{(n)}$ for $n=1,2,3,4$, and $\tF^{(5)}=\cF^{(5)}+\star_{10}\cF^{(5)}$. Our notation is such that $A\cdot B=A_{\mu\nu\cdots} B^{\mu\nu\cdots}$ and $A\cdot V\cdot B=A_{ab\cdots} B^{ab\cdots}$.

The nonlinear RR field strength however  is 
\beqa
F^{(n)}&=&\cF^{(n)}+H\wedge C^{(n-3)}
\eeqa
which is invariant under gauge transformation $\delta C= d\Lambda +H\wedge \Lambda$
where $C=\sum_{n=0}^8C^{(n)}$ and $\Lambda=\sum_{n=0}^7\Lambda^{(n)}$. So the RR gauge symmetry requires one to replace the linearized RR field strength $\cF^{(n)}$ in \reef{LTdual} and \reef{FF} by the nonlinear field strength $F^{(n)}$. This then  produces many new couplings. The above D-brane actions at order $\alpha'^2$, however, do  not include  all closed string couplings at this order, \eg the action does not include $H^4$ or $F^4$ couplings.  The action has also open string couplings.  In principle, all these couplings may be found by extracting the corresponding disk-level S-matrix elements. For example, the couplings involving   $H^4$   may be found by analyzing the disk-level S-matrix element of four vertex operators. However, it is very  hard to calculate such S-matrix element and to extract its four-momentum contact terms. See \cite{Stefanski:1998he,Garousi:2010bm,Garousi:2011ut,Becker:2011ar,BabaeiVelni:2016ahf} for the calculations of the disk-level S-matrix element of one specific  RR and two NSNS vertex operators from which some of the  couplings of one RR and two NSNS fields at order $\alpha'^2$ have been extracted. The disk-level S-matrix element of one arbitrary RR and two NSNS vertex operators have been calculated in \cite{Becker:2016bzb}. See   \cite{Fotopoulos:2001pt,Becker:2011ar,Garousi:2012gh,Jalali:2016xtv} for the S-matrix element of two massless closed and one open strings and also \cite{Fotopoulos:2001pt,Garousi:2000ea,Garousi:2007fk,Hatefi:2013mwa} for the calculations of disk-level S-matrix element of one RR and three open string states. One may use an extension of the KLT relation, to write the disk-level S-matrix elements of closed  and open string vertex operators   in terms of disk-level S-matrix elements  of only open string vertex operators \cite{Garousi:1996ad,Stieberger:2009hq,Stieberger:2015vya}. 

\subsection{$PR^2$-level amplitude of  two closed strings and its $\alpha'^2$-couplings}
 
 To introduce O$_p$-planes in type II theory, one may consider the $Z_2$ group defined by the transformations $\sigma\rightarrow -\sigma$ and $X^i\rightarrow -X^i$. The first one reverses the orientation of string and the second one is a reflection in spacetime. The orientation reversal interchanges  the right-moving   and  the left-moving modes. The type II theory has this global symmetry. If one gauges this  symmetry, then the states in the gauge invariant theory must be  invariant under this $Z_2$ transformations. For example, graviton which has no index along the  $i$-directions, survives because graviton is symmetric between right-moving and left-moving modes. Similarly, a Kalb-Ramond field with one index along the $i$-directions, survives because Kalb-Ramond is antisymmetric under interchanging the right-moving and left-moving modes. The extra minus sign resulted from the spacetime reflection then makes the Kalb-Ramond field to be invariant under the $Z_2$ transformations. The Kalb-Ramond state with no index along $i$-directions and graviton with one index along the $i$-directions are projected out. In the gauge invariant theory, there is an object, \ie O$_p$-plane,  orthogonal to the $i$-directions which causes the above projections. The O$_p$-plane is at the fixed point of the spacetime reflection, \ie it is  not dynamical objects.  As a result, there is no open string excitations for O$_p$-planes. However, they carry mass and charges so closed string fields can couple to the world-volume of O$_p$-planes. 

 In this section, we are going to review  the calculation of  the scattering amplitude of two massless closed strings  at projective-plane level in type II superstring theory and discuss how higher derivative couplings appear in the world-volume theory of O$_p$-planes.  
In type II theory    the  background charge of th projective plane  is $Q_{\phi}= 2$.   The S-matrix element \reef{AmpDO} for two    gravitons    at $PR^2$-level which represents the  scattering amplitude of one graviton off an O$_p$-plane, is
\beqa
A(1,2 )&\sim&g_s<V^{(0,0)}_1 V^{(-1,-1)}_2 >\labell{ampo}
\eeqa
The graviton vertex operator in the presence of O$_p$-plane has the following structure   \cite{Garousi:2006zh}:
\beqa
V(p,\veps)&=& \int d^2 z \bigg[\cV(p,p\inn D,\veps\inn D,z,\bz)+\cV(p,p\inn D,\veps\inn D,\bz,z) \bigg]\labell{ver3}
\eeqa
where the vertex operator $\cV(p,p\inn D,\veps\inn D,z,\bz)$ is the graviton vertex operator after using the doubling trick \reef{trick}.  The dimension of O$_p$-plane is specified by imposing the cross-cap  condition on world sheet fields $X^{\mu},\psi^{\mu}$. Because of the cross-cap, the   holomorphic and   antiholomorphi parts of fields on $PR^2$ are not independent. The propagators between holomorphic fields and between antiholomorphic fields are the same as the sphere propagators, however, the propagators between holomorphic and antiholomorphic fields, after using the doubling trick, are 
\beqa
<X^{\mu}(z)X^{\nu}(\bw)>&=&-\eta^{\mu\nu}\log(1+z\bw)\nn\\
<\psi^{\mu}(z)\psi^{\nu}(\bw)>&=&-\frac{\eta^{\mu\nu}}{1+z\bw}\nn\\
<\phi(z)\phi(\bw)>&=&- \log(1+z\bw)\labell{pro3}
\eeqa
 Unlike the disk calculation, here  one can not  map the amplitude to the upper-half plane, because the $RP^2$ has no boundary wherese the upper-half plane has boundary.  

Replacing the vertex operator \reef{ver3} into \reef{ampo}, and using the above propagators, one finds the integrand is invariant under the following transformations:
\beqa
z\rightarrow \frac{az+b}{cz+d}\,\,\,,\,\,\,\bz\rightarrow -\frac{d\bz-c}{b\bz-a}\,&;&ad-cb=1
\eeqa
which is consistent with the fact that the dimension of the moduli space of $PR^2$ with two punctures is one. Fixing this symmetry by setting $z_1=0$ and $|z_2|=r$, \ie
\beqa
\int d^2z_1d^2z_2&\rightarrow \int_0^1 dr^2
\eeqa
one can write again the result in terms of gamma functions. The result is \cite{Garousi:2006zh}
\beqa
A&\sim&\alpha'^2g_s^2T'_p\, K(1,2)\frac{\Ga(-t/2)\Ga(-u/2)}{\Ga(1-t/2-u/2)}\delta^{p+1}(p_1\inn V+p_2\inn V)\labell{ampo2}
\eeqa
where $T'_p$ is the tension of O$_p$-plane, 
$t=-\alpha'(p_1+p_2)^2/2$ and $u=-\alpha'(p_1+p_2\inn D)^2/2$. Using the fact that $(p_2\inn D)^{\mu}$ is transformation  of momentum $p_2^{\mu}$ under the $Z_2$  transformation, one observes that the two channels here are closed string channels. The kinematic factor $K(1,2)$ is exactly the same as for the D$_p$-branes case.  The  $PR^2$-level scattering amplitude of any other two massless closed string vertex operators  in type II supergravity has the same structure as \reef{ampo2}. The kinematic factor $K(1,2)$, however, depends on external states. In all cases the kinematic factors are the same as the corresponding factor for disk amplitude \cite{Garousi:2006zh}.

 One may expand the gamma functions in \reef{ampo2} around $\alpha'\rightarrow 0$, \ie
\beqa
A&\sim&\alpha'^2g_s^2T'_p\, K(1,2)\left(-\frac{4}{tu}-\frac{\pi^2}{6}+\cdots\right)\delta^{p+1}(p_1\inn V+p_2\inn V)\labell{ampo4}
\eeqa
where dots represent terms with higher orders   $\alpha'$.  The amplitude  has   simple poles at $u,t=0$ and    contact terms with infinite number of momenta. The simple poles are reproduced by the DBI action   and the supergravity. The higher derivative extension  of the DBI action should  produce the contact terms of the amplitude \reef{ampo4}. The amplitude \reef{ampo4} at order $\alpha'^2$ has only contact term, \ie
\beqa
A(\alpha'^2)&\sim&-\frac{\pi^2}{6}\alpha'^2g_s^2T'_p\, K(1,2) \delta^{p+1}(p_1\inn V+p_2\inn V)\labell{ampo44}
\eeqa
Since the kinematic factor for O$_p$-planes are exactly those for D$_p$-branes, the two closed string couplings   at order $\alpha'^2$ are exactly given by  \reef{RTN}, \reef{LTdual} and \reef{FF} in which the second fundamental form  is set to zero, \ie   geodesic embedding. In fact the gravity part of the second fundamental form \reef{sec} has one index along the $i$-directions, as a result, it is projected out under the $Z_2$ transformation. 
There are many couplings of massless closed string fields at order $\alpha'^2$ which do not appear in \reef{RTN}, \reef{LTdual} and \reef{FF}.   These couplings have also    genus and non-perturbative corrections. The latter can not be found from the perturbative S-matrix elements \reef{AmpDO}. So we have to use another technique to find such couplings. One expects the consistency of the couplings   in \reef{RTN}, \reef{LTdual} and \reef{FF}  with dualities may fix all other couplings.

\section{String dualities}

In this section, we are going to briefly review the T-duality and S-duality in string theory which lead to the conclusion that the five super string theories in 10-dimensional spacetime are not    fundamental, but are   different limits of the   M-theory. One may use these dualities to constrain the effective actions of the superstring theories and their non-perturbative objects at the higher order of $\alpha'$. We begin with the T-duality. See \cite{Giveon:1994fu,Alvarez:1994dn} for review articles on T-duality.

\subsection{T-duality}

If one  compares the spectrum of   the bosonic string theory on $R^{(25)}\times S^{(1)}$ where the  circle has radius $\rho$, and on $R^{(25)}\times \tS^{(1)}$ where the  circle has radius $\alpha'/\rho$, one would find the spectra in the two cases are identical, \ie the infinite tower of the winding modes in one case correspond to the infinite tower of the  Kaluza-Kelin modes in the other case, and vice versa. This indicates that the bosonic string theory is invariant under   T-duality when it is compactified on a circle.  Similarly, the type IIA  superstring theory  and the heteratic  $H_{\rm so(32)}$ theory on a circle with radius $\rho$ are T-dual of the type IIB and the heterotic $H_{E_8\times E_8}$  on a circle with radius $\alpha'/\rho$, respectively. 

If one extends the circle  to tours  $T^n$, then one would find that the spectrum of the  bosonic string theory is invariant under the T-duality group $O(n,n,Z)$. In fact, the mass spectrum of the free bosonic theory on $T^n$ with metric $G_{ij}$ and B-field $B_{ij}$ is (see \eg  \cite{Becker:2007zj})
\beqa
M^2&=& 4(N_R+N_L-2)+2(W^i\,\,\, K_i){\cG}^{-1}_{ij}\pmatrix{ W^j\cr K_j}
\eeqa
where $W^i,\, K_i$ are winding and Kaluza-Kelin numbers, $N_R,\, N_L$ are the right-moving and left-moving number operators, which satisfy the level-matching condition $N_R-N_L=W^iK_i$,  and ${\cG}_{ij}$ and its inverse are 
\beqa
{\cG}_{ij}=\pmatrix{ G^{-1}_{ij}&-G^{-1}_{ik}B_{kj}\cr B_{ik}G^{-1}_{kj} &G_{ij}-B_{ik}G^{-1}_{kl}B_{lj}  }&;&{\cG}^{-1}_{ij}=\pmatrix{G_{ij}-B_{ik}G^{-1}_{kl}B_{lj}&B_{ik}G^{-1}_{kj}\cr -G^{-1}_{ik}B_{kj}&G^{-1}_{ij}}\labell{genmet}
\eeqa
The spectrum and the level-matching condition are invariant under the following transformations:
\beqa
\left\{\matrix{W^i\rightarrow K_i\cr K_i\rightarrow W_i\cr \cG^{-1}\rightarrow \cG}\right.&;&\left\{\matrix{  K_i\rightarrow K_i+N_{ij}W^j\cr B_{ij}\rightarrow B_{ij}+N_{ij}}\right.\labell{trans}
\eeqa
where $N_{ij}$ is an antisymmetric matrix of integers. The first one is called inverse transformation and the second one is called shift transformation. 

Now consider the orthogonal group  $O(n,n,Z)$  whose  elements  satisfy the relation
\beqa
A^T\pmatrix{0&1_n\cr 1_n&0}A=\pmatrix{0&1_n\cr 1_n&0}\labell{ATA}
\eeqa
Under this group, the winding  and   the KK numbers   transform as doublet, \ie
\beqa
\pmatrix {W\cr K}\rightarrow O\pmatrix {W\cr K} 
\eeqa
and the generalized metric $\cG$ transform as 
\beqa
 \cG\rightarrow O\cG O^T
\eeqa  
The spectrum and the level-matching condition are invariant under these transformations. The two specific $O(n,n,Z)$ matrices 
\beqa
A=\pmatrix{0&1_n\cr 1_n&0}\,\,\,;\,\,A=\pmatrix{1_n&0\cr N_{ij}&1_n}
\eeqa
correspond to the inverse and to the shift transformations, respectively. They    generate an arbitrary $O(n,n,Z)$ matrix.

 When there is no winding and KK numbers, one may still consider the inverse transformation in \reef{trans} to relate the massless fields, \ie the metric and the B-field, on the tours to their corresponding fields in the  dual tours. One can write the inverse transformation on the generalized metric, \ie $\cG\rightarrow \cG^{-1}$ as the following transformation:
\beqa
Q_{ij}&\rightarrow &Q^{-1}_{ij}\labell{Qij}
\eeqa
where $Q_{ij}=G_{ij}+B_{ij}$. 

The above transformation should  be extended to  curved spacetime with background fields.   Such transformations   have been found by Buscher \cite{Buscher:1987sk,Buscher:1987qj}. To rederive them in the path-integral formalism \cite{delaOssa:1992vci}, consider   sphere-level path-integral \reef{Z3} in the bosonic string theory   that includes the  general background fields $G_{MN}(X^{\mu}, X^i)$,  $B_{MN}(X^{\mu}, X^i)$, and   $\phi(X^{\mu}, X^i) $, \ie
\beqa
Z&=&\int DX^M \,e^{-\int d^2 z \left(Q_{MN}\prt_z X^{M}\prt_{\bz}X^N +R^{(2)}\phi \right)}\labell{Z}
\eeqa
The   dilaton action   is one order of $\alpha'$   higher than the action for metric and B-field, \ie the dilaton is   one-loop action in the world-sheet theory. Now suppose the world-sheet action is invariant under global translation in $X^i$-directions, \ie $X^i\rightarrow X^i+\lambda^i$. This happens when the compact space is torus and the background fields are independent of $X^i$-directions, \ie $Q_{MN}=Q_{MN}(X^{\mu})$ and $\phi=\phi (X^{\mu})  $. This symmetry may be gauged by changing the ordinary derivatives in $X^i$-directions to the covariant derivatives, \ie $\prt_{\alpha} X^i\Rightarrow D_{\alpha} X^i=\prt_\alpha X^i+A_\alpha^i$, and then the measure of path-integral may be constrained by  the  delta function $\Delta [\frac{1}{2}\eps^{\alpha\beta}F_{\alpha\beta}^i]$ that imposes the gauge field strength to be  zero, \ie  the gauge field is pure gauge on the world-sheet with trivial topology. Introducing new fields $\tX^{i}$, one can write the delta function in path-integral form as  
\beqa
\Delta [\frac{1}{2}\eps^{\alpha\beta}F_{\alpha\beta}^i]&\sim&\int D\tX_n e^{-\frac{1}{2}\int d^2z \tX^i\eps^{\alpha\beta}F^i_{\alpha\beta}}
\eeqa
The gauge symmetry then make $Z$ to be infinite.   To have finite $Z$, one   may   fix the symmetry by fixing $X^i=0$ and dropping the volume of the gauge group. The path-integral then becomes
\beqa
 Z=\int DA D\bA DX'^M \,e^{-\int d^2 z \left(Q_{\mu\nu} \prt_z X^{\mu}\prt_{\bz}X^{\nu}+Q_{ij} A^{i}\bA^{j}+(Q_{\mu i} \prt X^{\mu}-\prt\tX^i)\bA^{i}+(Q_{i\nu} \bprt X^{\nu}+\bprt\tX^i)A^{i} +R^{(2)}\phi \right)}\nn
\eeqa
Integrating out the gauge fields, one would find dual theory in terms of dual coordinates $X'^M=(X^{\mu},\tX^i)$. Now using the following integral
\beqa
\int \prod_{k=1}^N dZ_kd\bZ_k\,e^{- (\bZ_k c_{kl}Z_l+a_kZ_k+b_k\bZ_k)}&\sim&\frac{1}{\det(c)}e^{a_k(c^{-1})_{kl}b_l}
\eeqa
one can perform the path-integral classically over the gauge fields to find the following result:
\beqa
Z&=&\int  DX'^{M} \,e^{-\int d^2 z \left(Q'_{MN} \prt_z X'^{M}\prt_{\bz}X'^{N} +R^{(2)}\phi' \right)}\labell{Z'}
\eeqa
where   $Q'_{MN}$ is
\beqa
Q'_{\mu\nu}&=&Q_{\mu\nu}-Q_{\mu i}(Q^{-1})^{ij}Q_{j\nu}\nn\\
Q'_{\mu i}&=&-Q_{\mu j} (Q^{-1})^j{}_i\nn\\
Q'_{i \mu}&=&(Q^{-1})_i{}^jQ_{j\mu}\nn\\
Q'_{ij}&=&Q^{-1}_{ij} \labell{Q'}
\eeqa
 The dilaton remains intact under the above classical calculations. Quantum mechanically, however, the Jacobin that comes from integrating out the gauge fields, produces corrections to the above transformations. At one loop-level, there is no correction to $Q'_{MN}$, however, the dilaton shifts as \cite{Buscher:1987sk,Buscher:1987qj}
\beqa
\phi'&=&\phi-\frac{1}{2}\ln\det(Q_{ij})\labell{phi'}
\eeqa
The transformations \reef{Q'} and \reef{phi'} are   extension of  the  transformation \reef{Qij} to the curved spacetime with background fields. The path-integral approach can easily be extended to the  superstring theories, which results the same  T-duality transformations  \reef{Q'} and \reef{phi'}. The path-integral approach   has been   used in \cite{delaOssa:1992vci} to study the T-duality transformations for the cases that the compact space has non-abelian isometries.

If the original theory \reef{Z} is conformal invariant, then the dual theory \reef{Z'} with \reef{Q'} and \reef{phi'} would be conformal invariant at one-loop level. One may impose the conformal invariance at higher-loop levels to find derivative corrections to \reef{Q'} and \reef{phi'}. On the other hand,   the conformal invariance of \reef{Z}   requires   vanishing of the world-sheet  beta functions which  would produce the equations of motion in spacetime. In other worlds, the invariance of $\beta=0$ under T-duality transformations is equivalent to the invariance of the spacetime equations of motion under T-duality. This invariance may  in turn be implemented in the spcetime effective actions that produce the equations of motion\footnote{It has been observed in \cite{Haagensen:1997er,Olsen:1998yw} that the renormalization group flows, \ie the beta functions, at one-loop level are also invariant under the Buscher rules, and at two-loop level are invariant under the Buscher rules plus their corrections at order $\alpha'$ which have been found in  \cite{Tseytlin:1991wr}.}. One may impose this constraint on the effective actions   not only to find the derivative corrections to the transformations \reef{Q'} and \reef{phi'}, but also to find constraints on the higher derivative terms of the  effective actions.  
   
The T-duality transformations of the non-perturbative objects D$_p$-branes/O$_p$-planes depend on whether they are along or orthogonal to the $X^i$-directions along which the T-duality are imposed. If a D$_p$-brane/O$_p$-plane is along the $X^i$-directions,  it transforms to D$_{p-n}$-brane/O$_{p-n}$-plane orthogonal to the $X^i$-directions in the T-dual theory, and vice versa.  For D$_p$-brane,  the T-duality changes the Newman boundary conditions along the $X^i$-directions to the  Dirichlet boundary conditions in the T-dual theory which in turn changes the gauge fields along the $X^i$-directions     to the transverse scalar fields in the T-dual theory, \ie
\beqa
A_i&\rightarrow &\chi^i\labell{A'}
\eeqa
 The T-duality relates the brane tensions as $V^{(n)}T_p=T_{p-n}$ where $V^{(n)}$ is the volume of torus $T^n$. Their world-volume effective actions should satisfy the corresponding duality. That is, the T-duality of the effective action of a D$_p$-brane/O$_p$-plane along the $X^i$-directions  should be equivalent to the world-volume effective action of  D$_{p-n}$-brane/O$_{p-n}$-plane orthogonal to the $X^i$-directions. 

The effective action of D$_p$-brane  at the leading order of $\alpha'$ in type II superstring theory is given by the DBI and WZ actions. The DBI action, in the absence of the massless open string fields, for D$_p$-brane along $T^n$  and D$_{p-n}$-brane arthogonal to $T^n$ are  
\beqa
S^{\rm DBI}_p=-T_p\int d^{p+1}x\,e^{-\phi}\sqrt{-\det(Q_{ab})}&;&S^{\rm DBI}_{p-n}=-T_{p-n}\int d^{p-n+1}x\,e^{-\phi}\sqrt{-\det(Q_{\ha\hb})}\labell{DBI}
\eeqa
where $a,b$ are the world volume indices of the D$_p$-brane, and $\ha,\hb$ are the world volume indices of the D$_{p-n}$-brane. Now if   the fields in $S^{\rm DBI}_p$ are independent of the tours coordinates $X^i$, then one   uses dimensional reduction along the tours $T^n$ and then uses the T-duality, \ie
\beqa
 S^{\rm DBI}_p=-T_{p-n}\int d^{p-n+1}x\,e^{-\phi'}\sqrt{-\det(Q'_{ab})}
\eeqa
Using the relations \reef{Q'} and $\reef{phi'}$, one finds the above action is in fact the DBI action \reef{DBI} for D$_{p-n}$-brane. 

The T-duality transformations on the RR fields in type II superstring theories have been found in \cite{Meessen:1998qm} by requiring the solutions of type IIA supergravity to be transformed under the T-duality to the solutions of type IIB supergravity. These transformations may be rederived by using the  fact that the D$_p$-brane effective action should be transformed to  the D$_{p-n}$-brane effective action.  The WZ action, in the absence of the massless open string fields, for D$_p$-brane along $T^n$ and D$_{p-n}$-brane orthogonal to $T^n$are  
\beqa
S^{\rm WZ}_p=-T_p\int_{M^{p+1}} e^B C&;&S^{\rm WZ}_{p-n}=-T_{p-n}\int_{M^{p-n+1}} e^B C
\eeqa
where $C=\sum_{n=0}^8C^{(n)}$ and one should consider $p+1$-forms in $S^{\rm WZ}_p$ and $p-n+1$-forms in $S^{\rm WZ}_{p-n}$. Now if   the fields in $S^{\rm WZ}_p$ are independent of the tours coordinates $X^i$, then one   uses dimensional reduction along the tours $T^n$ and then uses the T-duality, \ie
\beqa
 S^{\rm WZ}_p=-T_{p-n}\int_{M^{p-n+1}} e^{B'} C'\labell{WZ'}
\eeqa
  Now if one uses the following transformation:
\beqa
( e^{B'} C')_{\ha_1\ha_2\cdots\ha_{p-n+1}i_1i_2\cdots i_n}&=&( e^{B} C)_{\ha_1\ha_2\cdots\ha_{p-n+1}}\labell{C1}
\eeqa
where $i_1,\cdots, i_n$ are indices along the $T^n$, then the action \reef{WZ'} would be the WZ action for D$_{p-n}$-brane. If one considers a D$_{p-n}$ brane orthogonal to the tours $T^n$, then after  T-duality  the WZ action transforms to the corresponding term in D$_p$-brane action along $T^n$ provided that
\beqa
( e^{B'} C')_{\ha_1\ha_2\cdots\ha_{p-n+1}}&=&( e^{B} C)_{\ha_1\ha_2\cdots\ha_{p-n+1}i_1i_2\cdots i_n}\labell{C2}
\eeqa
One may use the T-duality transformation \reef{Q'} for B-field on \reef{C1} and \reef{C2} to find the T-duality transformation for the RR potentials. When there is only one killing direction $y$,  the transformation is \cite{Meessen:1998qm}
\beqa
C'^{(n)}_{\mu\cdots \nu \alpha y}&=& C^{(n-1)}_{\mu\cdots \nu \alpha}-(n-1)\frac{C^{(n-1)}_{[\mu\cdots\nu|y}G_{|\alpha]y}}{G_{yy}}\labell{C'}\\
C'^{(n)}_{\mu\cdots\nu\alpha\beta}&=&C^{(n+1)}_{\mu\cdots\nu\alpha\beta y}+nC^{(n-1)}_{[\mu\cdots\nu\alpha}B_{\beta]y}+n(n-1)\frac{C^{(n-1)}_{[\mu\cdots\nu|y}B_{|\alpha|y}G_{|\beta]y}}{G_{yy}}\nn
\eeqa
Using the transformations \reef{C1} and \reef{C2}, one may extend the T-duality transformation \reef{C'} to the cases that there is more than one killing direction. The compatibility of  the DBI action, in the presence of the abelian massless open string fields, with T-duality  have been observed in \cite{Myers:1999ps}. The T-duality invariance of the    WZ action, in the presence of nonabelian massless open string fields, has been used in \cite{Myers:1999ps} to find the Myers term. They have been confirmed with explicit S-matrix calculations in \cite{Garousi:2000ea}.

The D$_p$-brane/O$_p$-plane effective actions at the leading order of $\alpha'$   are then manifestly invariant under the T-duality  transformations \reef{Q'}, \reef{phi'}, \reef{A'} and \reef{C'}\footnote{In our convention, the transverse scalar fields of D$_p$-brane, $\chi^i$, have the same dimension as $X^{\mu}$. The T-duality transformation \reef{A'} then indicates that in our convention the gauge field, $A_a$, has also the same dimension. As a result, the DBI action in the presence of gauge field is at order $\alpha'^0$. }.  There is also a manifestly T-duality invariant action for type II supergravities which is in terms of  $C=\sum_{n=0}^8C^{(n)}$  \cite{Fukuma:1999jt}. This indicates that not only the effective actions at the leading order of $\alpha'$ are invariant under the T-duality transformations, but also  the presence of brane does not change the form of the transformation rules. One expects the effective actions at the higher order of $\alpha'$ to be also invariant under the T-duality.   

The transformation rules \reef{Q'}, \reef{phi'}, \reef{A'} and \reef{C'}, however,  may receive higher derivative corrections.
Suppose  the T-duality operator has an $\alpha'$ expansion
\beqa
T&=&\sum_{n=0}^{\infty}(\alpha')^nT^{(n)}\,,
\eeqa
where $T^{(0)}$ is given by  \reef{Q'}, \reef{phi'}, \reef{A'} and \reef{C'}. The invariance of the effective actions \reef{eff} under the T-duality transformation, \ie
\beqa
S_{\rm eff}&\stackrel{T }{\longrightarrow}&S_{\rm eff}
\eeqa
then means the action at the leading   order of  $\alpha'$ to be invariant under the leading term of the $T$-operator, \ie 
\beqa
S_0&\stackrel{T^{(0)}}{\longrightarrow}&S_0\,.
\eeqa
At order $\alpha'$, the action has two terms, \ie $S=S_0+\alpha' S_1$. The invariance then means
\beqa
S_1&\stackrel{T^{(0)}}{\longrightarrow}&S_1+\delta S\,,\nonumber\\
S_0&\stackrel{T^{(1)}}{\longrightarrow}&-\delta S\,.
\eeqa
At order $(\alpha')^2$, the action has three terms, \ie $S=S_0+\alpha' S_1+(\alpha')^2 S_2$ and again the invariance means that
\beqa
S_2&\stackrel{T^{(0)}}{\longrightarrow}&S_2+\delta S_1+\delta S_2\,,\nonumber\\
S_1&\stackrel{T^{(1)}}{\longrightarrow}&-\delta S_1\,,\nonumber\\
S_0&\stackrel{T^{(2)}}{\longrightarrow}&-\delta S_2\,.
\eeqa
Similarly for the action at higher orders of $\alpha'$.

By studying the effective actions of the bosonic and heterotic string theories  at order $\alpha'$, it has been shown in \cite{Kaloper:1997ux} that the transformations \reef{Q'}, \reef{phi'} do receive higher derivative corrections at order $\alpha'$, \ie $T^{(1)}\ne 0$. It has been observed  in \cite{Garousi:2013gea} that the same T-duality transformations   are required to show that the D$_p$-brane action at order $\alpha'$  in the bosonic theory is invariant under T-duality. We will review this calculation in section 6.

In  above approach, it has been assumed   the effective action \reef{eff} to be  invariant under the general coordinate  and the B-field gauge transformations. The invariance under the T-duality then requires the T-duality transformations to receive $\alpha'$-corrections. One may release the general covariance and the invariance under the B-field gauge transformation in the effective action \reef{eff}, but requires it to be invariant under $T^{(0)}$-transformations, \ie
\beqa
S_{\rm eff}&\stackrel{T^{(0)} }{\longrightarrow}&S_{\rm eff}
\eeqa
The above constraint has been used in \cite{Garousi:2015qgr} to find  non-covariant corrections at order $\alpha'$   to the D$_p$-brane effective action  in the bosonic string theory.  We will review this calculation in section 6. A systematic approach for constructing the  non-covariant effective actions, however,  
 is the Double Field Theory  (DFT) \cite{Hull:2009mi,Hohm:2010jy} in which the generalized metric $\cG_{ij}$ \reef{genmet} is used and the actions are required to be explicitly invariant under
$O(D,D,R)$ transformations where $D$ is the dimension of spacetime. The modification of this theory to Double $\alpha'$-geometry in which the generalized
Lie derivative receives $\alpha'$-corrections, requires and determines the higher derivative couplings \cite{Hohm:2013jaa,Marques:2015vua}.  

Another non-covariant approach for constructing the $\alpha'$-corrections in manifestly $O(n,n,R)$ invariant form is to reduce the theory on $T^n$ and observe that  the scalar fields, \ie the scalar fields that appear in the  generalized metric ${\cG}_{ij}$, satisfy the   relation \reef{ATA} and ${\cG}_{ij}$ is symmetric matrix \cite{Godazgar:2013bja}. This metric has an $\alpha'$-expansion, \ie
\beqa
\cG&=&\cG_0+\alpha'\cG_1+\alpha'^2\cG_2+\cdots
\eeqa
where $\cG_0$ is the one in \reef{genmet}, $\cG_1$ constructed from the second partial derivatives of scalars $(G_{ij},B_{ij})$,  $\cG_2$ constructed from the forth partial derivatives of the scalars $(G_{ij},B_{ij})$ and so on. Constraining $\cG$ to be symmetric and satisfy \reef{ATA}, one may find $\cG_1,\, \cG_2,\cdots$. This constraint on the scalar fields may fix the form of unreduced action \cite{Godazgar:2013bja}. We are not interested in this approach and in the DFT approach in this review article. 

We are interested in the simple case that  the theory is compactified on a circle with the killing coordinate $y$ and radius $\rho$. In this case, the Buscher rules \reef{Q'} and \reef{phi'}     become
\beqa
e^{2\phi'}=\frac{e^{2\phi}}{G_{yy}}&;& 
G'_{yy}=\frac{1}{G_{yy}}\nonumber\\
G'_{\mu y}=\frac{B_{\mu y}}{G_{yy}}&;&
G'_{\mu\nu}=G_{\mu\nu}-\frac{G_{\mu y}G_{\nu y}-B_{\mu y}B_{\nu y}}{G_{yy}}\nonumber\\
B'_{\mu y}=\frac{G_{\mu y}}{G_{yy}}&;&
B'_{\mu\nu}=B_{\mu\nu}-\frac{B_{\mu y}G_{\nu y}-G_{\mu y}B_{\nu y}}{G_{yy}}\labell{nonlinear}
\eeqa
where $\mu,\nu$ denote any   direction other than $y$. In above transformation the metric is  in the string frame.  
One may be interested in  studying the S-matrix elements under the above T-duality transformations. In the S-matrix elements, the vertex operators correspond to small perturbations of fields around the flat background. Assuming that the  massless  fields are small perturbations around the background, \ie
\beqa
G_{\mu\nu}&=&\eta_{\mu\nu}+  h_{\mu\nu}\,;\,\,G_{yy}\,=\,\frac{\rho^2}{\alpha'}(1+  h_{yy})\,;\,\,
\phi\,=\,\phi_0+  \Phi
\eeqa
 then the  nonlinear  transformations \reef{nonlinear} and \reef{C'}  take the following linear form for the perturbations:
\beqa
&&
  \Phi'= \Phi - \frac{1}{2}h_{yy},\,h'_{yy}=-h_{yy},\, h'_{\mu y}=B_{\mu y}/\eta_{yy},\, B'_{\mu y}=h_{\mu y}/\eta_{yy},\,h'_{\mu\nu}=h_{\mu\nu},\,B'_{\mu\nu}=B_{\mu\nu}\nonumber\\
&&C'^{(n)}_{\mu\cdots \nu y}={ C}^{(n-1)}_{\mu\cdots \nu },\,\,\,C'^{(n)}_{\mu\cdots\nu}={ C}^{(n+1)}_{\mu\cdots\nu y}\labell{linear}
\eeqa
where $\eta_{yy}=\rho^2/\alpha'$. The above linear transformations are used in section 4 for studying T-duality Ward identity. One may also use them, in some cases,  to study the invariance of a subset of the couplings  in the effective actions at a given order of $\alpha'$ under linear T-duality.

To study the invariance of the full effective action under T-duality, however, one must use the nonlinear transformations \reef{nonlinear} and \reef{C'}. In that case, it is convenient to use the following dimensional reduction on  the 10-dimensional metric and Kalb-Ramond field:
  \beqa
G_{MN}=\left(\matrix{g_{\mu\nu}+e^{\varphi}g_{\mu }g_{\nu }& e^{\varphi}g_{\mu }&\cr e^{\varphi}g_{\nu }&e^{\varphi}&}\right)\,,\qquad B_{MN}= \left(\matrix{b_{\mu\nu}+\frac{1}{2}b_{\mu }g_{\nu }- \frac{1}{2}b_{\nu }g_{\mu }&b_{\mu }\cr - b_{\nu }&0&}\right)\labell{reduc}\eeqa
where $g_{\mu\nu}, \,b_{\mu\nu}$ are the metric and the B-field, and $g_{\mu},\, b_{\mu}$ are two vectors  in the 9-dimensional base space. Inverse of the 10-dimensional metric is 
\beqa
G^{MN}=\left(\matrix{g^{\mu\nu} &  -g^{\mu }&\cr -g^{\nu }&e^{-\varphi}+g_{\alpha}g^{\alpha}&}\right)\labell{inver}
\eeqa
where $g^{\mu\nu}$ is inverse of the 9-diemsional metric which raises the index of the   vectors.  In this parametrization, the 9-dimensional dilaton is  $\bar{\phi}=\phi-\varphi/4$. The T-duality transformations \reef{nonlinear} in this parametrization simplify to the following linear transformations:
\beqa
\varphi'= -\varphi
\,\,\,,\,\,g'_{\mu }= b_{\mu }\,\,\,,\,\, b'_{\mu }= g_{\mu } \labell{T2}
\eeqa
The 9-dimensional base space fields $g_{\alpha\beta}$, $b_{\alpha\beta}$ and $\bar{\phi}$ remain invariant under the T-duality. The T-duality of the RR fields \reef{C'} become
\beqa
C'^{(n)}_{\mu\cdots \nu \alpha y}&=& C^{(n-1)}_{\mu\cdots \nu \alpha}-(n-1)C^{(n-1)}_{[\mu\cdots\nu|y}g_{\alpha]} \labell{C'1}\\
C'^{(n)}_{\mu\cdots\nu\alpha\beta}&=&C^{(n+1)}_{\mu\cdots\nu\alpha\beta y}+nC^{(n-1)}_{[\mu\cdots\nu\alpha}b_{\beta]}+n(n-1) C^{(n-1)}_{[\mu\cdots\nu|y}b_{\alpha}g_{\beta]} \nn
\eeqa
which remains nonlinear. 

In the covariant approach, the transformations \reef{T2} and \reef{C'1} in general   receive $\alpha'$ corrections. However, as we have seen in the previous section, the higher derivative corrections to the type II supergravities starts at order $\alpha'^3$. That means the $\alpha'$ corrections to the above T-duality transformations in type II superstring theory starts at order $\alpha'^3$.  On the other hand, the first corrections to the D$_p$-brane/O$_p$-plane effective action starts at order $\alpha'^2$. So one can use the   T-duality transformation  \reef{T2} and \reef{C'1} to study the brane actions at order $\alpha'^2$. We expect the compatibility of the brane couplings in \reef{RTN}, \reef{LTdual} and \reef{FF} with the above T-duality transformations and S-duality transformations that we are going to review in the next subsection, enables one to finds all couplings at order $\alpha'^2$.  

\subsection{S-duality}

The careful  studies  of  the 10-dimensional supergravities which are the low-energy effective actions of the superstring theories,   and the 11-dimensional supergravity which is the low energy effective action of M-theory, reveals  that there is  a $Z_2$ transformation that relates type I supergravity at couplings $g_s$ to the H$_{\rm SO(32)}$-supergravity at coupling $g_s^{-1}$, the type IIA at couplings $g_s$ to the dimensional reduction of the 11-dimensional supergravity on a circle with radius $g_s$,  the H$_{E_8\times E_8}$-supergravity at couplings $g_s$, to the dimensional reduction of the 11-dimensional supergravity on a line with length $g_s$, and  the type IIB at couplings $g_s$ to the type IIB at couplings $g_s^{-1}$. The last   transformation  is in fact promoted to the  $SL(2,R)$ transformation. The $Z_2$ symmetries are expected to be the symmetries of the corresponding superstring theories/M-theory, and the $SL(2,Z)$ subgroup of $SL(2,R)$,   is expected to be the symmetry of type IIB superstring theory. Since the weak coupling constant transforms to the strong coupling constant in these duality transformations, they are called  S-duality.  We are interested only in the S-duality of the type IIB superstring theory. See \eg \cite{Schwarz:1996bh} for a review article on the S-duality.

    Under  the $SL(2,R)$ transformations, the  graviton in  the Einstein frame, \ie  $G^E_{\mu\nu}=e^{-\phi/2}G_{\mu\nu}$   and the RR four-form are  invariant. The B-field and the  RR two-form transform as doublets \cite{ Gibbons:1995ap,Tseytlin:1996it,Green:1996qg}:
\beqa 
\cB\ \equiv\ 
\pmatrix{B \cr 
C^{(2)}}\rightarrow (\Lambda^{-1})^T \pmatrix{B \cr 
C^{(2)}} 
\eeqa
where the matrix $\Lambda=\pmatrix{d&c\cr b&a} \in SL(2,R)$.    The transformation of the dilaton and the RR scalar $C$ appears in the transformation of   the  $SL(2,R)$ matrix $\cM$ 
\beqa
 {\cal M}=e^{\phi}\pmatrix{|\tau|^2\ C \cr 
C\ 1}\label{M}
\eeqa
where  $\tau=C+ie^{-\phi}$. This matrix  transforms as \cite{ Gibbons:1995ap}
\beqa
{\cal M}\rightarrow \Lambda {\cal M}\Lambda ^T\labell{TM}
\eeqa
For the special case that $C=0$, and for  the particular $SL(2,R)$ matrix $\cN=\pmatrix{0&1\cr -1&0}$, one finds the weak-strong transformation $e^{-\phi}\rightarrow e^{\phi}$. 

The manifestly $SL(2,R)$-invariant form of the type IIB supergravity  \reef{sugraB} is 
\beqa
  S_{IIB}&\supset& \frac{1}{2\kappa^2}\int d^{10}x  \sqrt{-G}\bigg[R+\frac{1}{4}\Tr(\nabla_{\mu}\cM \nabla^{\mu}\cM^{-1})-\frac{1}{12}\cH^T_{\mu\nu\rho}\cM\cH^{\mu\nu\rho}-\frac{1}{4}|F_{(5)}|^2\bigg]\nn\\
	&&\qquad\quad\qquad\qquad\qquad\qquad\quad\qquad\qquad\qquad\qquad\qquad-\frac{1}{8\kappa^2}\int C_{(4)} \cH^T\cN \cH\labell{IIB1}
\eeqa
where $\cH=d\cB$ and  the five-form field strength is $F_{(5)}=dC_{(4)}+\frac{1}{2}\cB^T\cN\cH$.
A similar   expression is expected for corrections at all higher order of $\alpha'$. 

One may expect the $\alpha'$-corrections   to involve  only $\cM$, $\cH$, $F^{(5)}$, $C^{(4)}$ and metric which transform as tensors under the $SL(2,R)$. However, unlike the two derivative action, the higher derivative actions have both genus and non-perturbative contributions as well. So the action should involve some $SL(2,Z)$ tensors representing these contributions. Such tensor for $\alpha'^3$-corrections in which we are interested, have been found in \cite{Green:1997tv}. Consider the gravity couplings in \reef{Y3}. In the Einstein frame and for constant dilaton, they are
\beqa
S_{II}\supset \frac{\gamma \z(3)}{3.2^7\kappa^2} \int d^{10}x e^{-3\phi/2} \sqrt{-G}(t_8t_8R^4+\frac{1}{4}\eps_{8}\eps_{8} R^4)\labell{Y31}
\eeqa
The dilaton factor indicates that the above action is not invariant under the $SL(2,R)$ transformation so there are some missing terms.  This is consistent with the fact that the above action does not include the genus and the non-perturbative contributions. The   $SL(2,Z)$ invariant form of the action \reef{Y31} has been conjectured in  \cite{Green:1997tv}  to be
\beqa
S_{IIB}\supset\frac{\gamma }{3.2^8 }\int d^{10}x E_{(3/2)}(\tau,\bar{\tau}) \sqrt{-G}(t_8t_8R^4+\frac{1}{4}\eps_{8}\eps_{8}R^4)\labell{Y2}
\eeqa
where $E_{(3/2)}(\tau,\bar{\tau}) $ is the $SL(2,Z)$ invariant non-holomorphic Eisenstein series. For general $s$, the $SL(2,Z)$ invariant function  $E_{(s)}(\tau,\bar{\tau}) $  is defined as 
\beqa
E_{(s)}(\tau,\bar{\tau})&=&\sum_{(n,m)\ne(0,0)}\frac{\tau_2^s}{|m+n\tau|^{2s}}
\eeqa
where $\tau_1+i\tau_2=\tau$. It   satisfies the following eigenvalue  equation:
\beqa
 \tau_2^2\prt_{\tau}\prt_{\bar{\tau}}E_{(s)}&=&s(s-1)E_{(s)} 
\eeqa
which   has two solutions $e^{-s\phi}$ and $e^{-(1-s)\phi}$ corresponding to two particular orders of perturbation theory, and infinite number of non-perturbative solutions.
$E_{(3/2)}(\tau,\bar{\tau})$ has the following weak-expansion \cite{Green:1997tv}:
\beqa
E_{(3/2)}(\tau,\bar{\tau})
&\!\!\!\!=\!\!\!\!&2\z(3)\tau_2^{3/2}+4\z(2)\tau_2^{-1/2}+8\pi\tau_2^{1/2}\sum_{m\neq 0,n\geq 1}\left|\frac{m}{n}\right| K_{1}(2\pi|mn|\tau_2)e^{2\pi imn\tau_1}\labell{series}
\eeqa
 where $K_1$ is the Bessel function.  The above expansion shows that there are no perturbative corrections beyond the tree level and one-loop level, but there are an infinite number  of D-instanton  corrections.  By explicit calculation, it has been shown in \cite{DHoker:2005vch} that there is no two-loop correction to the action \reef{Y2}. The modular invariant function $E_{(3/2)}$ should appear for all NSNS and RR couplings at order $\alpha'^3$. Apart from this overall factor, all couplings should be combined appropriately to be written in $SL(2,R)$ invariant form as in type IIB supergravity \reef{IIB1}.

Since the RR four-form is invariant under the $SL(2,R)$ transformations, the effective action of   O$_3$-plane should be  invariant under the S-duality. The effective action at the leading order of $\alpha'$ in the Einstein frame is 
\beqa
S_{O_3}&\supset&-T'_3\int d^4 x \sqrt{-\det(\tG_{ab})}-T'_3\int \tC_4
\eeqa
where the tilde-sign means pull-back operator, \eg $\tG_{ab}=\prt_a X^{\mu}\prt_bX^{\nu} G_{\mu\nu}$. In the static gauge, \ie $X^a=\sigma^a, X^i=0$, one finds $\tG_{ab}=G_{ab}$ and $\tC_{abcd}=C_{abcd}$. This action is obviously invariant under the S-duality. 

There is similar symmetry for D$_3$-brane action at the leading order. However, the D-brane effective action contains the gauge field $A_a$ and the transverse scalar fields $\chi^i$ in the static gauge, \ie $X^a=\sigma^a, X^i=\chi^i$. The gauge symmetry requires also the gauge field strength and the Kalb-Ramond potential appear in the effective action as $\tB+F$. The transverse scalars appear in the action through the pull-back operator and through the dependence of the closed string fields on the transverse coordinates \cite{Garousi:1998fg}. The action at the leading order of $\alpha'$ is
\beqa
S_{D_3}\supset-T_3\int d^4 x \sqrt{-\det(\tG_{ab}+\tB_{ab}+F_{ab})}-T_3\int [\tC_4+(\tB+F)\tC_2+\frac{1}{2}(\tB+F)^2C_0]\labell{D3}
\eeqa
 The transverse scalar fields are invariant under the S-duality, and the gauge field transforms as \cite{Gibbons:1995ap}
 \beqa
\cF=\pmatrix{*F \cr 
G_F}\rightarrow (\Lambda^{-1})^T \pmatrix{*F \cr 
G_F} \labell{F'}
\eeqa
where the antisymmetric tensor $(G_F)_{ab}$ is defined in terms of the Lagrangian as 
\beqa
(G_F)_{ab}&=&-\frac{2}{T_3}\frac{\prt L}{\prt F^{ab}}
\eeqa
Obviously, because of the presence of B-field in the DBI part, the   action \reef{D3} is not invariant under the S-duality. However, the equations of motion are invariant under the S-duality \cite{Gibbons:1995ap,Green:1996qg}\footnote{One may consider the string excitation of the D$_3$-brane to be a $(p,q)$-string. In that case, one considers two gauge fields that transform as doublet under the $SL(2,R)$ transformation. Then one can write $SL(2,R)$-covariant action which includes both gauge fields \cite{Bergshoeff:2006gs}. We are, however, interested in  the case that only F$_1$-string propagates one the world-volume of D-branes.}.     If one ignores the couplings which include $\tB+F$, then the action \reef{D3} is invariant under the S-duality.
 One expects the higher derivative couplings in  the O$_3$-plane   theory and the higher derivative couplings in  the D$_3$-brane theory, except the couplings involving  $\tB+F$   to be invariant under the S-duality. The higher derivative couplings involving $\tB+F$, on the other hand,  are expected to be invariant at the equations of motion level. 

The S-duality   requires, among other things, that the tree-level couplings to be extended to include the higher genus couplings. Using the KLT relations, and the fact that the eight-derivative couplings in the bulk action \reef{Y2} include $E_{(3/2)}$ which has only one-loop corrections,  one expects the world-volume four derivative couplings to have also corrections only at one-loop level. In other word, the world-volume action should include the Eisenstein series $E_{(1)}$ which has only tree and one-loop contributions at the weak-coupling expansion. For $s=1$, however, the series \reef{series} diverges logarithmically. The regularized function which is proportional to the modular invariant function $\log(\tau_2|\eta(\tau)|^4)$, has  the following weak-expansion \cite{Bachas:1999um,Basu:2008gt}:
\beqa
E_{(1)}(\tau,\bar{\tau})&=&\z(2)\tau_2-\frac{\pi}{2}\ln(\tau_2)+\pi\sqrt{\tau_2}\sum_{m\neq 0,n\neq 0}\left|\frac{m}{n}\right|^{1/2}K_{1/2}(2\pi|mn|\tau_2)e^{2\pi imn\tau_1}\labell{regE}
\eeqa 
The first term is tree-level contribution and the second terms is one-loop contribution. The modular invariant function $E_{(1)}$ should appear for almost all NS, NSNS and RR couplings at order $\alpha'^2$. Apart from this overall factor, the couplings should be combined appropriately to be written in $SL(2,R)$ invariant form.

The  Eisenstein series $E_{(1)}$, however, should not appear for the world-volume couplings at order $\alpha'^2$  which have been found from the anomaly cancellation mechanism because they have no genus contribution at all. These couplings should involve a modular function which has only one perturbative contribution and  infinite number of D-instanton contributions. The curvature squared corrections to the WZ action \reef{4-der}  which have been found from the anomaly cancellation mechanism,  has one RR scalar field $C$ which is not invariant under the global $SL(2,Z)$. This term has been extended to   the anomalous modular function $\log({\eta(\tau)/\bet(\btau)})$   in \cite{Bachas:1999um,HenryLabordere:2001zv} which produces $C$ and infinite number of D-instanton  contributions at the weak expansion.   The $\tau$-dependent anomalous transformation for the case of D$_3$-brane in trivial normal bundle, cancels the  $\tau$-dependent   anomalous transformation of the Jacobean of the  massless modes of the D$_3$-brane, and a $\tau$-independent modular anomaly remains \cite{Bachas:1999um}.

Apart from the anomalous couplings \reef{4-der}, all other world-volume couplings at order $\alpha'^2$ should have the overall factor of $e^{-\phi}$ in the Einstein frame  which must be extended to $E_{(1)}$. Even the couplings that are related to \reef{4-der} by the T-duality transformation have the overall factor of $e^{-\phi}$ \cite{Garousi:2011fc}. That means the couplings which are related by the T-duality to the anomalous couplings are not anomalous. So such couplings can not be found by the anomaly cancellation mechanism. 

We expect the consistency of the bulk  couplings \reef{Y3} and the brane couplings  \reef{4-der}, \reef{RTN} with the duality transformations enable one to find all spacetime couplings at order $\alpha'^3$ and all the world-volume couplings at order $\alpha'^2$. The duality transformations can also appear in the S-matrix elements as the duality Ward identities which may be used to generate the S-matrix elements. In the next section we review the duality Ward identities. 

\section{Duality Ward identities as generating functions  }   

All S-matrix elements of any   gauge theory   satisfy     Ward identity which is invariance of the S-matrix elements under 
    linear gauge transformations on the quantum fluctuations and the full nonlinear transformations on the background fields.  
 Similar Ward identities exist for almost all S-matrix elements under the global duality transformations. The S-matrix elements corresponding to the anomalous couplings, however, do not satisfy the duality Ward identities. Since the  duality transformations are global, the momenta in the S-matrix elements are invariant under the duality transformations. The background fields in the S-matrix element should transform according to the duality transformations in the previous sections, and the polarization tensors should transform according to the linearized form of the duality transformations, \eg \reef{linear}.  The linear dualities may transform one field to some other fields, as a result, they may transform one S-matrix element to some other  S-matrix elements. This means the duality  Ward identities may generate some S-matrix element from a given S-matrix element. 

To clarify it,  suppose,  using the prescription \reef{AmpDO}, one calculates an S-matrix element at tree-level  in the flat spacetime with constant  dilaton background $\phi_0$ and finds the following result\footnote{ We have normalized the vertex operators in the amplitude \reef{AmpDO} with the factor of $g_s$ for each closed string vertex operator, and with $\sqrt{g_s} $ for each open string vertex operator. However, if one is going to correspond the vertex operators to the supergraviton fields, \eg $B, C^{(2)}$,   then the normalization factors make inconsistency  because $g_s$ is not invariant under the duality transformations. So in order to study the amplitude \reef{AmpDO} under the duality, one should either assume the $g_s$ corresponding to the vertex operators is inert, or one should normalize the vertex operators without $g_s$ and $\sqrt{g_s}$ factors, \ie drop these factors from the amplitude \reef{AmpDO}. We will use this latter assumption.}:
\beqa
A_{\rm tree}&\sim&K_1(\z_i,p_i)f_1(s,t,u,\cdots)+K_2(\z_i,p_i)f_2(s,t,u,\cdots)  +\cdots\labell{AS}
\eeqa
 where $K_1, K_2,\cdots$ are some kinematic factors, and $f_1, f_2, \cdots$ are some functions of the Mandelstam variables that represent the poles of the amplitude. The flat metric in the Mandelstam variables and in the kinematic factors is the string frame metric. 

If one is going to study this amplitude under the S-duality Ward identity, the amplitude should be written in the Einstein frame, \ie 
\beqa
A_{\rm tree}&\sim&K_1(\z_i,p_i,\phi_0)f_1(s,t,u,\cdots,\phi_0)+K_2(\z_i,p_i,\phi_0)f_2(s,t,u,\cdots,\phi_0)  +\cdots\labell{AE}
\eeqa
where $\phi_0$ results from transforming the metric to $e^{\phi_0/2}\eta_{\mu\nu}$. To extend the amplitude to satisfy the S-duality Ward identity, one should first include the constant background RR scalar $C_0$ into the amplitude because the background dilaton trsnsforms to $C_0$ under the S-duality. This constant field should be added to the factors $f_1,f_2,\cdots$ in such a way to make them invariant under the $SL(2,Z)$ transformations. The constant field $C_0$ and some other supergraviton fields should be also added to the kinematic factors to make them invariant under the linear $SL(2,R)$ transformations, \ie
\beqa
A^s\sim K^s_1(\z_i, \veps_i,p_i,\phi_0,C_0)f^s_1(s,t,u,\cdots,\phi_0,C_0)+K^s_2(\z_i,\veps_i,p_i,\phi_0,C_0)f^s_2(s,t,u,\cdots,\phi_0,C_0)  +\cdots\nn
\eeqa
where $\veps_i$ are the polarization tensor of the other  supergraviton fields. The new amplitude $A^s$ is assumed to satisfy the S-duality Ward identity, \eg it contains the appropriate loop-level amplitudes and non-perturbative effects. Now one may set $C_0$ in $f_1, f_2, \cdots$ to   zero to find tree-level amplitude, \ie
\beqa
A_{\rm tree}^s\sim K^s_1(\z_i,\veps_i,p_i,\phi_0,C_0)f_1(s,t,u,\cdots,\phi_0)+K^s_2(\z_i,\veps_i,p_i,\phi_0,C_0)f_2(s,t,u,\cdots,\phi_0)  +\cdots\nn
\eeqa
The above  amplitude is now  the sum of the amplitude \reef{AE} and some other tree-level amplitudes in the Einstein frame which the S-duality Ward identity generates them.

As an example, consider the disk-level S-matrix element of one dilaton, one Kalb-Ramond and one gauge boson vertex operators on the world-volume of D$_p$-brane. The result for D$_3$-brane in the Einstein frame is \cite{Garousi:2012gh}
 \beqa
 A _{\rm tree}&\sim&T_3\phi_1e^{-\phi_0}F^{ab}H_{\mu ba}f^{\mu}\labell{A1}
 \eeqa
 where  $\phi_1$ is polarization of dilaton, $H$ is the field strength of the polarization of the Kalb-Ramond field and $F$ is the field strength of the polarization of the gauge boson. In the above amplitude $f^{\mu}$ is
\beqa
f^{\mu}&=&e^{-\phi_0/2}\left(I_{11}\bigg[\frac{p_1.D.p_1}{p_1.p_2}  p_1.V^{\mu}+\frac{p_1.k_3\,p_1.D.p_1}{(p_1.p_2)^2}p_1^{\mu}\bigg]\right.\nn\\
 &&\left.-I_2\bigg[4p_1.V^{\mu}+\frac{p_2.D.p_2}{p_1.p_2} \left(2 p_1^{\mu}- p_1.N^{\mu}\right)+\frac{p_1.k_3\,p_2.D.p_2}{(p_1.p_2)^2}p_1^{\mu}\bigg]\right)\nn
\eeqa
where $I_2,I_{11}$ are the functions of the Mandelstam variables representing the poles of the amplitude \cite{Garousi:2012gh}. The function $f^{\mu}$ at the leading order of $\alpha'$ which corresponds to the supergravity and the D-brane  action at the leading order of $\alpha'$,  has no dilaton factor \cite{Garousi:2012gh}. The $SL(2,R)$-extension of the amplitude   \reef{A1} is
\beqa
 A^s _{\rm tree}&\sim&T_3 (*\cF_l)^{Tab}\delta\cM_1\cH_{\mu ba}f^{\mu}\labell{A6}
 \eeqa
where $\cF_l$ is the linearized form of $\cF$ in \reef{F'}, \ie $G=e^{\phi_0}F-C_0(*F)$, $\cH=d\cB$ and $\delta\cM_1$ is the variation of $\cM$ around the background constant fields $\phi_0,C_0$, \ie
\beqa
\delta\cM_1=\pmatrix{-(e^{-\phi_0}-C_0^2e^{\phi_0}) \phi_1+2C_0 e^{\phi_0}  C_1& C_0 e^{\phi_0} \phi_1+e^{\phi_0}  C_1\cr 
C_0 e^{\phi_0} \phi_1+e^{\phi_0}  C_1&e^{\phi_0}  \phi_1}\labell{dM}
\eeqa
where  $C_1$ is polarization of the RR scalar field.
 It transforms under the $SL(2,R)$ the same as matrix $\cM$ transforms in \reef{TM}.
 The $SL(2,R)$ invariant expression $(*\cF_l)^T \delta\cM_1\cH $   has the following six components \cite{Garousi:2012gh}\footnote{There is a type  in the exponential factors in the  last two terms in eq.(30) in \cite{Garousi:2012gh}. The dilaton factor must be $e^{\phi_0}$.}:  
\beqa
(*\cF)^T\delta\cM_1\cH&=&e^{-\phi_0}\phi_1FH+\phi_1(*F)F^{(3)}+C_0\phi_1(*F)H\nonumber\\
&&+C_1(*F)H-e^{\phi_0}C_0C_1FH-e^{\phi_0}C_1FF^{(3)}\labell{dual}
\eeqa
 where $F^{(3)}$ is the field strength of the polarization of the RR two-form. Therefore, the amplitude \reef{A6} represents six different S-matrix elements at the disk-level. It has been shown in \cite{Garousi:2012gh} that the explicit calculation confirms the amplitudes involving $\phi_1(*F)F^{(3)}$ and $C_1(*F)H$. The amplitude involving $e^{\phi_0}C_1FF^{(3)}$ is also consistent with explicit calculation \cite{GM}. The $SL(2,R)$-extension of the disk-level S-matrix element of one graviton, one B-field and one gauge boson on the world-volume of D$_3$-brane has been studied in \cite{Garousi:2012gh}, and the $SL(2,R)$-form of one closed and three non-abelian open strings has been studied in \cite{Garousi:2011jh,Garousi:2011we}. One may also use the S-duality Ward identity to find the S-matrix elements on the world-volume of NS$_5$-brane and F$_1$-string \cite{Garousi:2011we} in type IIB theory.

The anomalous coupling \reef{4-der} on the world-volume of D$_3$-brane is not invariant under the S-duality even in the presence of background field $C_0$. As a result, the S-matrix element of one RR scalar and two graviton vertex operators does not satisfy the S-duality Ward identity. In the bulk, however, there is no anomalous coupling. So all S-matrix elements should have $SL(2,R)$-extension. For example, the $SL(2,R)$-extension of the sphere-level S-matrix element of two gravitons and two Kalb-Ramond vertex operators at eight-momentum level in spacetime \reef{Y3} can be written as 
\beqa
A_{\rm tree}^s&\sim&e^{-3\phi_0/2}t_8t_8 RR(D\cH)^T\cM_0 D\cH\labell{AH}
\eeqa
where $D\cH_{abcd}=\cH_{ab[c,d]}$. This contains four different amplitudes, \ie
\beqa
 D\cH^T\cM_0D\cH =e^{-\phi_0}(1+e^{2\phi_0}C_0^2)DHDH+e^{\phi_0}DFDF-e^{\phi_0}C_0(DHDF+DFDH)\labell{SHH}
\eeqa
The S-matrix element of two gravitons and two RR two-forms in \reef{AH} is confirmed by explicit calculation \cite{Bakhtiarizadeh:2013zia}.

If one is going to  study an amplitude   under the T-duality Ward identity, one should consider the amplitude in the string frame \reef{AS}. The factors  $f_1, f_2, \cdots$ are invariant under the T-duality because there is no dilaton in them in the string frame. Assuming the killing direction is $y$, one should first use dimensional reduction which separates the indices of the polarization tensors in the kinematic factors to $y$ and $\mu\ne y$, \ie
 \beqa
A_{\rm tree}&\sim&K_1(\z^{\mu}_i,\z^y_i,p^{\mu}_i)f_1(s,t,u,\cdots)+K_2(\z^{\mu}_i,\z^y_i,p^{\mu}_i))f_2(s,t,u,\cdots)  +\cdots\nn
\eeqa
Then one should transform them under the linear T-duality transformations \reef{linear}, \ie
\beqa
A^t_{\rm tree}&\sim&K_1(\z^{'\mu}_i,\z'^y_i,p^{\mu}_i)f_1(s,t,u,\cdots)+K_2(\z'^{\mu}_i,\z'^y_i,p^{\mu}_i))f_2(s,t,u,\cdots)  +\cdots\nn
\eeqa
This generates the dimensional reduction of a new amplitude for the T-dual fields. In this case, unlike the S-duality Ward identity,     the  reduced   amplitude does not fully fix the form of the amplitude because there may be couplings which vanishes after   the dimensional reduction. However, imposing other constraints as well like the gauge symmetry Ward identity or S-duality Ward identity may fix the form of the new amplitude. 

Consider, for example, the S-matrix element of two gravitons and two RR two-form vertex operators in spacetime coordinates at order $\alpha'^3$ in the string frame, 
\beqa
A_{\rm tree} &\sim& t_8t_8 RRDF^{(3)}DF^{(3)}\nn\\
&\sim& \frac{1}{16} F_{r s[p,q]}^2 R_{h k m n}^2+\frac{1}{4} F_{r s[m,n]} F_{r s[p,q]} R_{h k m n} R_{h k p q}\nonumber\\&&+F_{h s[p,q]} F_{r s[p,q]} R_{h k m n} R_{k r m n}-2 F_{h s[q,m]} F_{r s[p,q]} R_{h k m n} R_{k r n p}-2 F_{h s[p,q]} F_{r s[q,m]} R_{h k m n} R_{k r n p}\nonumber\\&&+F_{h s[p,q]} F_{r s[m,n]} R_{h k m n} R_{k r p q}+F_{k r[m,n]} F_{r s[p,q]} R_{h k m n} R_{h s p q}+\frac{1}{2} F_{k r[p,q]} F_{h s[p,q]} R_{h k m n} R_{r s m n}\nonumber\\&&+\frac{1}{8} F_{h k[m,n]} F_{r s[p,q]} R_{h k p q} R_{r s m n}+4 F_{k r[n,p]} F_{h s[p,q]} R_{h k m n} R_{r s m q}+F_{k r[m,n]} F_{h s[p,q]} R_{h k m n} R_{r s p q}\nonumber\\&&-F_{k r[n,p]} F_{h s[q,m]} R_{h k m n} R_{r s p q} 
\eeqa
 Then use the dimensional reduction and consider the terms that the $y$-index appears in the RR fields. Under linear T-duality \reef{linear} the $y$-index drops. So it produces the amplitude of two gravitons and two RR one-forms at order $\alpha'^3$, \ie
\beqa
A^t_{\rm tree} &\sim& \frac{1}{4} F_{p q,r}^2 R_{h k m n}^2+\frac{1}{2} F_{m n,s} F_{p q,s} R_{h k m n} R_{h k p q}+F_{p q,h} F_{p q,r} R_{h k m n} R_{m n k r}\nonumber\\&&+2 F_{h s,q} F_{r s,q} R_{h k m n} R_{m n k r}+F_{k r,q} F_{h s,q} R_{h k m n} R_{m n r s}+2 F_{m q,r} F_{p q,h} R_{h k m n} R_{n p k r}\nonumber\\&&+2 F_{m q,h} F_{p q,r} R_{h k m n} R_{n p k r}+2 F_{h s,p} F_{r s,m} R_{h k m n} R_{n p k r}+2 F_{h s,m} F_{r s,p} R_{h k m n} R_{n p k r}\nonumber\\&&-4 F_{k r,n} F_{h s,q} R_{h k m n} R_{m q r s}+F_{m n,r} F_{p q,h} R_{h k m n} R_{p q k r}-F_{m n,k} F_{p q,s} R_{h k m n} R_{p q h s}\labell{F2R}
\eeqa
They are reproduced by explicit S-matrix calculation \cite{Bakhtiarizadeh:2013zia}. Similarly one can find the S-matrix element of two gravitons and two arbitrary RR potentials \cite{Garousi:2013nfw} . Using T-duality and S-duality Ward identities, all four-point S-matrix elements of NSNS and RR vertex operators have been found and confirmed by explicit calculations \cite{Garousi:2013nfw,Bakhtiarizadeh:2013zia,Bakhtiarizadeh:2015exa}.

The disk-level S-matrix element of two closed string vertex operators at four-momentum level in spacetime are given by the couplings \reef{RTN}, \reef{LTdual} and \reef{FF}. It has been shown in \cite{Garousi:2009dj,Garousi:2010ki,Garousi:2011fc} that these couplings    satisfy the duality Ward identities, \ie the couplings of two RR two-form and the couplings of two Kalb-Ramond can be written as \cite{Garousi:2011fc}
\beqa
A^s_{\rm tree}&\sim&  \frac{1}{6}{\cal H}^T_{ijk,a}\cM\cH^{ijk,a}+\frac{1}{3}\cH^T_{abc,i}\cM\cH^{abc,i}-\frac{1}{2}\cH^T_{bci,a}\cM\cH^{bci,a}
\labell{FF2}
\eeqa
 which satisfies the S-duality Ward identity. The   S-matrix element of one RR $(p-3)$-form , one NSNS and one NS vertex operators on the world volume of D$_p$-brane has been calculated in \cite{Becker:2011ar,Garousi:2012gh}. The T-duality Ward identity on this amplitude has been used in \cite{Velni:2013jha} to generate the S-matrix elements for the RR $(p-1)$-form, $(p+1)$-form and RR $(p+3)$-form. These S-matrix elements are confirmed by explicit calculations in \cite{Jalali:2016xtv}.   The   S-matrix element of one RR $(p-3)$-form and two  NSNS   vertex operators  has been calculated in \cite{Becker:2011ar,Garousi:2011ut}. The T-duality Ward identity on this amplitude has been used in \cite{Velni:2013jha} to generate the S-matrix elements for arbitrary  RR form. They should be consistent with the corresponding S-matrix elements from explicit calculations \cite{Becker:2016bzb}.   

Having found the S-matrix elements either explicitly or by using the duality Ward identities which is complicated  for higher n-point functions, one may then study them at low energy to find the massless poles and the contact terms. They  should be reproduced by effective actions. The massless poles of the disk-level S-matrix element of two closed strings and the massless poles of the sphere-level S-matrix element of four closed strings at low energy are at the leading order of $\alpha'$ and the contact terms of these amplitudes are the higher order of $\alpha'$. This makes it easy to find the contact terms at order $\alpha'^2$ in the disk-level and the contact terms at order $\alpha'^3$  in the sphere-level. In general, however, there are  massless poles and contact terms  at the same order of $\alpha'$ so one has to carefully reproduce the massless poles by the effective field theory before interpreting the  contact terms as   new couplings of the effective theory because the difference between the massless poles of S-matrix element and the massless poles of the field theory may be some contact terms. This makes it difficult to extract new couplings at a given order of $\alpha'$   from the contact terms of the corresponding S-matrix elements. So one may impose the  T-duality and S-duality constraints directly on the effective actions. In this case, however, one has to use the full nonlinear duality transformations, \eg \reef{nonlinear} or \reef{TM}, on the fields in the effective action. In the next section, we use the T-duality constraint on the effective action of O$_p$-plane to find all NSNS couplings at order $\alpha'^2$ including the couplings with structure $H^4$ \cite{Robbins:2014ara,Garousi:2014oya}. These couplings  should be reproduced by  the $PR^2$-level S-matrix element of four B-field vertex operators which is a very hard calculation.
  
\section{T-duality constraint on O-plane action at order $\alpha'^2$} 

We have seen that the T-duality transformations in type II superstring theories receive no $\alpha'$-correction at order $\alpha'^2$ for the massless NSNS and RR  fields. So one should be able to find a world-volume covariant action for these fields at order $\alpha'^2$ which is consistent with the standard T-duality transformations \reef{nonlinear} and  \reef{C'}. This may not be the case if one includes the massless NS fields at this order because the transformation \reef{A'} may receive $\alpha'$-corrections.  To simplify the calculations, we consider only the world-volume NSNS couplings. On the other hand, the B-field appear in the world-volume theory either as field strength $H$ or as potential in the combination $F+\tB$. The latter form, however, is not $\alpha'$-dependent. So any order of this field may appear in the world-volume action at order $\alpha'^2$. We expect the  T-duality constraint would fix the presence of these terms. To simplify further, we consider the NSNS couplings on the world volume of O$_p$-plane which has no $(F+\tB)$-term at all.

We have seen in section 2.4 that two NSNS couplings at order $\alpha'^2$ on the world-volume of O$_p$-plane are given by the couplings \reef{RTN} in which the second fundamental form is zero. At this order there are also couplings with structure $RH^2$, $R(\nabla\phi)^2$, $\nabla^2\phi(\nabla\phi)^2$, $(\nabla\phi)^4$, $H^2 (\nabla\phi)^2$, $H^2\nabla^2\phi$ and $H^4$. There are two metrics for contracting the spacetime indices of these bulk tensors. One is the first fundamental form, \ie $\tG^{\mu\nu}=\prt_a X^{\mu}\prt_b X^{\nu}\tG^{ab}$
which   projects bulk indices to the world-volume. The second one is $
\bot^{\mu\nu}=G^{\mu\nu}-\tG^{\mu\nu} $
which projects the bulk indices to the transverse space. One can contract the bulk indices with $(\tG^{\mu\nu}, \bot^{\mu\nu})$,  with $(G^{\mu\nu}, \bot^{\mu\nu})$ or with $(G^{\mu\nu}, \tG^{\mu\nu})$. We use the last pair for contracting the indices. Using the package ``x-Act``\cite{Nutma:2013zea}, one can write all such contractions with unknown coefficients. We assume the coefficients for the NSNS couplings to be independent of the dimension of O$_p$-plane, as the couplings in \reef{RTN}.

The O$_p$-plane  couplings should be invariant under the $Z_2$ transformations $\sigma\rightarrow -\sigma$ and $X^i\rightarrow -X^i$. This projects out $\nabla\cdots\nabla B$ with even number of world-volumes indices,  $\nabla\cdots\nabla \phi$ and  $\nabla\cdots\nabla R$ with odd number of world-volumes indices. So after writing all the contractions, one should separate the spacetime indices to world volume and transverse indices. Then the coefficients should be constraint such that in the couplings there would be none of the above terms.

To find the T-duality constraint on the coefficients, one needs the reduction of the couplings. The dimensional reduction of $G^{\mu\nu}$ in the parametrization \reef{reduc} is given in \reef{inver}. The dimensional reduction of $\tG^{\mu\nu}$, however, depends on weather the O$_p$-plane is along or orthogonal to the circle. When O$_p$-plane is along the circle, one finds in the static gauge where $X^{a}=\sigma^a,\, X^i=0$, the pull-back of metric and the first fundamental form become \cite{Garousi:2014oya}
\beqa
\tG_{ab}=\pmatrix{g_{\ta\tb}+e^{\varphi}g_{\ta}g_{\tb}&e^{\varphi}g_{\ta}\cr e^{\varphi}g_{\tb}&e^{\varphi} }&;&\tG^{\mu\nu}=\pmatrix{g^{\ta\tb}&-g^{\ta}\cr-g^{\tb}&e^{-\varphi}+g_{\ta}g^{\ta}}
\eeqa
where $\ta,\tb$ are the world-volumes indices which do not include the world-volume direction $y$. When O$_{p-1}$-plane is orthogonal to the circle, one finds that the reduction of the pull-back of metric and the first fundamental form are \cite{Garousi:2014oya} 
\beqa
\tG_{ab}=\pmatrix{g_{\ta\tb}&0\cr0&0}&;&\tG^{\mu\nu}=\pmatrix{g^{\ta\tb}&0\cr0&0}
\eeqa
Using the above reductions, one observes that the reduction of O$_p$-plane action at order $\alpha'^0$ when it is along the circle is 
\beqa
\int d^{p+1} x e^{-\phi}\sqrt{-\tG_{ab}}\rightarrow \int d^p x e^{-\phi+\varphi/2}\sqrt{-\det g_{\ta\tb}}\labell{dba1}
\eeqa
On the other hand, the reduction of O$_{p-1}$-plane action at order $\alpha'^0$ when it is orthogonal to  the circle is 
\beqa
\int d^{p+1} xe^{-\phi}\sqrt{-\tG_{ab}}\rightarrow  \int d^p x e^{-\phi}\sqrt{-\det g_{\ta\tb}}\labell{dba2}
\eeqa
Obviously the   transformation of \reef{dba1} under the T-duality rule \reef{T2} is identical to \reef{dba2}. The same thing should happen  for all $\alpha'^2$ couplings. That is, the T-duality of the reduction of the world-volume action of O$_p$-plane when it is along the circle which we call it  $S^{wT}$, should be equal to the reduction of the world-volume action of O$_{p-1}$-plane when it is orthogonal to the circle which we call it  $S^t$. Therefore, the T-duality constraint is
\beqa
\int d^p x e^{-\phi} \sqrt{-\det g_{\ta\tb}}(\cL^{wT}-\cL^t)&=&0
\eeqa
In imposing the above constraint, one should drop the terms that are total derivatives and then set the coefficients of independent terms to be zero. This together with the $Z_2$ projection fix the unknown coefficients to be \cite{Robbins:2014ara,Garousi:2014oya}
 \beqa
S_p^{DBI} &\!\!\!\!\supset\!\!\!\!&\frac{\pi^2\alpha'^2T'_{p}}{48}\int d^{p+1}xe^{-\phi}\sqrt{-\tG}\bigg[H^{abi} H_{a}{}^c{}_i \cR_{bc} -\frac{3}{2} H^{abi} H_{ab}{}^j \cR_{ i j}+\frac{1}{2} H^{ijk} H_{ij}{}^l\cR_{ k l}\nonumber\\&&-H^{abi} H^{cd}{}_i R_{abcd}+H^{abi} H_i{}^{jk} R_{abjk}-\frac{1}{4} H^{abi} H_{ab}{}^j H_i{}^{kl} H_{jkl}+\frac{1}{4} H^{abi} H_{ab}{}^j H^{cd}{}_i H_{cdj}\nonumber\\&& +\frac{1}{8} H^{abi} H_{a}{}^{cj} H_b{}^d{}_j H_{cdi}-\frac{1}{6} H^{abi} H_{a}{}^{cj} H_{bc}{}^k H_{ijk}+\frac{1}{24} H^{ijk} H_i{}^{lm} H_{jl}{}^n H_{kmn}\bigg]\labell{finalH}
\eeqa
where $\cR$ is the same as $\bar{\cR} $ in \reef{Rij} in which the second fundamental form is zero. These couplings together with the couplings in \reef{RTN} give all NSNS couplings on the world-volume of O$_p$-plane at order $\alpha'^2$. Since these couplings are invariant under the standard Buscher rules \reef{Q'}, using the generalized metric $\cG$ in \reef{genmet}, one may be able to rewrite them in manifestly $O(D,D,R)$ invariant form as in the DFT formalism. 

We have found the above couplings which have no RR field, by requiring the couplings \reef{RTN} to be consistent with full T-duality transformations. One may use the consistency of the couplings \reef{LTdual} with the T-duality transformations to find all couplings at order $\alpha'^2$ which include one RR field. These couplings may  also include the standard WZ term at order $\alpha'^2$, \ie  \reef{4-der}. Similarly one may use the consistency of the couplings \reef{FF} with the T-duality to find all couplings at order $\alpha'^2$ which include two RR fields. Some of these couplings may be related to \reef{finalH} by the $SL(2,R)$ transformation. The S-duality transformation of \reef{finalH} produces also couplings which have four RR two-forms. The subsequent T-duality transformations  may fix all RR couplings at order $\alpha'^2$. 

\section{ T-duality constraint on D-brane  action   at order $\alpha'$}

The effective action of D-brane in bosonic string theory includes various world volume couplings of open string tachyon, transverse scalar fields, gauge field, closed string tachyon, graviton, dilaton and B-field. Duo to the presence of the   tachyons, the bosonic string theory and its D-branes   are all unstable. Assuming the tachyons are freezes at the top of their potentials, \ie the tachyon fields are zero, the effective action of the D$_p$-brane at the leading order of $\alpha'$ is given by the DBI action which is invariant under the T-duality transformation \reef{nonlinear} and \reef{A'}. The first higher derivative correction to this action in the  bosonic string theory is at order $\alpha'$. As a result, the first higher derivative corrections to the T-duality transformations \reef{T2} and \reef{A'} are at order $\alpha'$. Such corrections for the closed string fields \reef{T2} have been found in \cite{Kaloper:1997ux}. 
They are
\beqa
\varphi&\stackrel{T}\longrightarrow &-\varphi- \alpha'\lambda_0 \bigg[2(\nabla\varphi)^2+e^{\varphi}V_{\mu\nu}V^{\mu\nu}+e^{-\varphi}W_{\mu\nu}
W^{\mu\nu}\bigg]\,,\nonumber\\
g_{\mu}&\stackrel{T}\longrightarrow &b_{\mu}-\alpha'\lambda_0\,\bigg[2W_{\mu\nu}\nabla^{\nu}\varphi+e^{\varphi}H_{\mu\nu\lambda}V^{\nu\lambda}\bigg]\,,\nonumber\\
b_{\mu}&\stackrel{T}\longrightarrow &g_{\mu}-\,\alpha'\lambda_0\,\,\bigg[2V_{\mu\nu}\nabla^{\nu}\varphi-e^{-\varphi}H_{\mu\nu\lambda}W^{\nu\lambda}\bigg]\,, \nonumber\\
H_{\mu\nu\lambda}&\stackrel{T}\longrightarrow &H_{\mu\nu\lambda}-12\alpha'\lambda_0\bigg[\nabla_{[\mu}(W_{\nu}{}^{\rho}V_{\lambda]\rho})+\frac{1}{2}V_{[\mu\nu}W_{\lambda]\rho}\nabla^2\varphi
+\frac{1}{2}W_{[\mu\nu}V_{\lambda]\rho}\nabla^2\varphi\,, \nonumber\\
&&\qquad\,\,\,+\frac{1}{4}e^{\varphi}V^{\rho\chi}V_{[\mu\nu}H_{\lambda]\rho\chi}-\frac{1}{4}e^{-\varphi}W^{\rho\chi}W_{[\mu\nu}H_{\lambda]\rho\chi}\bigg]\,.\labell{corrT}
\eeqa
and the metric $g_{\mu\nu}$ and $\bar{\phi}$ remain invariant. In above transformations, $H$ is the field strength of the two form $b_{\mu\nu}$, \ie $H_{\mu\nu\lambda}=\prt_{\mu}b_{\nu\lambda}+\prt_{\lambda}b_{\mu\nu }+\prt_{\nu}b_{\lambda\mu}$, $V_{\mu\nu}$ is the field strength of $g_{\mu}$, \ie $V_{\mu\nu}=\prt_{\mu}g_{\nu}-\prt_{\nu}g_{\mu}$ and $W_{\mu\nu}$ is the field strength of $b_{\mu}$, \ie $W_{\mu\nu}=\prt_{\mu}b_{\nu}-\prt_{\nu}b_{\mu}$. The constant $\lambda_0$ is -1/4 for the bosonic string theory, is -1/8 for the heterotic string theory and is zero for the superstring theory.
Using these corrections, one may be able to find a covariant action for D$_p$-brane/O$_p$-plane which includes only the massless closed string fields. 

The covariant D$_p$-brane action at order $\alpha'$ should involve $R, \Omega,\,\nabla\phi, H$ and also B-field potential where both indices are   the world-volume indices. Since B-field is dimensionless, any order of $B_{ab}$ may appear in the D-brane action at order $\alpha'$. To simplify the calculation, we consider only second order of  fields. Writing all such couplings involving spacetime curvature, the second fundamental form, dilaton and the B-field at order $\alpha'$ and constraining the couplings to be consistent with the T-duality transformations   \reef{corrT},  at the second order of fields, one finds the following result \cite{Garousi:2013gea}:
\beqa
S^{DBI}_p  \!\!\!&\supset&\!\!\!
-\frac{ \alpha'T_p}{2}\int d^{p+1}xe^{-\phi}
\sqrt{-\tG }\bigg[\tR+2\bot_{\mu\nu}(\Omega_{ a}{}^a{}^{\mu}
\Omega_{ b}{}^b{}^{\nu}-{\Omega_{ab}{}^{\mu}
\Omega^{ab}{}^{\nu}})  +2\bot_{\mu\nu} 
\Omega_a{}^a{}^{\mu}\prt^{\nu}\phi   + \prt_{\mu}\phi\prt^{\mu}\phi \nonumber\\
&&-\frac{1}{8}\tilde{H}^2- \frac{1}{8} \bot^{\mu\nu}
H^2_{\mu\nu}   +\frac{1}{8} \bot^{\alpha\beta} \bot^{\mu\nu}
H_{\alpha\mu\lambda}H_{\beta\nu}{}^{\lambda} + \frac{1}{24} \bot^{\alpha\beta} \bot^{\mu\nu}\bot^{\lambda\sigma} H_{\alpha\mu\lambda}H_{\beta\nu\sigma}  \bigg]\labell{dbi2}
\end{eqnarray}
where $\tilde{H}^2=\tG^{\mu\nu}\tG^{\alpha\beta}\tG^{\rho\sigma}H_{\mu\alpha\rho}H_{\nu\beta\sigma}$ and $\tR=\tG^{\mu\nu}\tG^{\alpha\beta}R_{\mu\alpha\nu\beta}$. This action is consistent with the disk-level S-matrix element of two massless closed string vertex operators in the bosonic string theory \cite{Corley:2001hg,Garousi:2013gea}. Each term in the above action should  be multiplied by a function of B-field potential. See \cite{Ardalan:2002qt}, for a non-covariant form for such functions in the gravity part which have been found from the disk-level S-matrix element of two graviton vertex operators in the presence of constant B-field in the bosonic string theory. One may find covariant  functions by requiring the above couplings to be invariant under full non-linear T-duality transformations \reef{corrT}. Such functions for the O$_p$-plane, however, is trivial as the $Z_2$ transformation projects out the  couplings in which B-field has even number of world-volume indices. This projection on the above action produces the following action:
\beqa
S^{DBI}_p  \!\!\!&\supset&\!\!\!
-\frac{ \alpha'T'_p}{2}\int d^{p+1}xe^{-\phi}
\sqrt{-\tG }\bigg[\tR    + \prt_{a}\phi\prt^{a}\phi - \frac{1}{8} H_{abi}H^{abi}
  + \frac{1}{24} H_{ijk}H^{ijk} \bigg] \labell{boso}
\end{eqnarray}
which includes all couplings at order $\alpha'$ and should be  invariant under full T-duality transformation \reef{corrT}.

The $\alpha'$ corrections to the T-duality transformation of the open string fields \reef{A'}  have not been found yet.   One may consider a covariant action for massless open  string fields at order $\alpha'$ which includes all contactions of the second fundamental form $K_{ab}^{i}$ \reef{sec}, the gauge field strength $F_{ab}$ and its covariant derivative $D_aF_{bc}$ with the pull-back metric. The second fundamental form couplings have been already found from the covariant action in terms of the  massless closed string fields \reef{dbi2}. In our convention that the gauge field strength $F_{ab}$ and the brane velocity $\prt_a\chi^i$ are  dimensionless, the covariant action has four-field, six-field, and higher order couplings at order $\alpha'$. At four-field level,
imposing the covariant action to be consistent with the T-duality transformation \reef{A'}, one finds a bunch of couplings which are not consistent with the S-matrix element of four open string vertex operators in the bosonic string theory \cite{Garousi:2015qgr}. This is what one expects because the T-duality transformation \reef{A'} should receive $\alpha'$-correction in the covariant approach to the T-duality.  One may add some $\alpha'$-correction to \reef{A'} and constrain   the T-duality invariant couplings to be consistent with the S-matrix elements. In this way one may  find the $\alpha'$-corrections to the leading T-duality transformation \reef{A'} as well as  the effective action of four massless open string fields  at order  $\alpha'$. Similar calculation may lead one to find the six-field, the eight-field and all higher order couplings.

Alternatively, one may use the  non-covariant approach and consider all contractions of $F_{ab}$, $\prt_a F_{bc}$, $\prt_a\chi^i$ and $\prt_a\prt_b\chi^i$  at order $\alpha'$ with flat metric and constrain the non-covariant action to be consistent with the T-duality transformation \reef{A'}, with the second fundamental form couplings in \reef{dbi2}, and with the four-point  S-matrix elements. In this way, even though the action is not covariant, however,   one is able to find the couplings to all orders of $F$ and $\prt\chi$. Ignoring the T-duality invariant couplings which are total derivative terms and excluding the T-duality invariant couplings  which have  $dF$, one finds  \cite{Garousi:2015qgr}
 \beqa
S^{DBI}_p&\!\!\!\! \supset\!\!\!\!&	-T_p\int d^{p+1}\sigma\sqrt{-\det(\teta_{ab}+F_{ab} )}\bigg[1+\alpha'\bigg( 
 G'^ {ab} G'^ {cd}  \bot'^{ij}  \omega_ {abi} \omega_ {cdj}- 
 G'^ {ac} G'^ {bd}  \bot'^{ij}  \omega_ {abi} \omega_ {cdj}  \nonumber\\&&\qquad\qquad\qquad
-\frac{1}{2} G'^ {ad} G'^ {be} G'^ {cf}\psi_ {abc}\psi_ {def} - 
 G'^ {ab} G'^ {ce} G'^ {df}\psi_ {abc}\psi_ {def}+ 
 \frac{2}{3} G'^ {cf} \Theta^ {ad} \Theta^ {be}\psi_ 
{abc}\psi_ {def}\nonumber\\&&\qquad\qquad\qquad  - 
\frac{8}{3} G'^ {bc} G'^ {de} \Theta^ {ai}\psi_ {bcd} \omega_ 
{aei} + G'^ {cd} \Theta^ {ai} \Theta^ {bj} \omega_ 
{acj} \omega_ {bdi} + 
 \frac{4}{3} G'^ {bj} \Theta^ {ai} \Theta^ {cd} \omega_ 
{acj} \omega_ {bdi} \nonumber\\&&\qquad\qquad\qquad+ 
 \frac{5}{3} G'^ {cd} \Theta^ {ai} \Theta^ {bj} \omega_ 
{aci} \omega_ {bdj} - 
 \frac{8}{3} G'^ {cd} \Theta^ {ai} \Theta^ {bj} \omega_ 
{abi} \omega_ {cdj} + 
 \frac{4}{3}  \bot'^{ij} \Theta^ {ac} \Theta^ {bd} 
\omega_ {abi} \omega_ {cdj}  \nonumber\\&&\qquad\qquad\qquad  + 
 2 G'^ {bd} G'^ {ce} \Theta^ {ai}\psi_ {bac} \omega_ 
{dei}- 2 G'^ {bc} G'^ {de} \Theta^ {ai}\psi_ {bac} 
\omega_ {dei}   - 
 \frac{2}{3} G'^ {ai} G'^ {ce} \Theta^ {bd}\psi_ {bac} \omega_ 
{dei}\nonumber\\&&\qquad\qquad\qquad - 2 \Theta^ {ai} \Theta^ {bd} 
\Theta^ {ce}\psi_ {bac} \omega_ {dei}\bigg)+O(\alpha'^2)\bigg]\labell{final}
\eeqa
where $\psi_{abc}=\prt_aF_{bc}$, $\omega_{abi}=\prt_a\prt_b\chi_i$, $\bot'^{\mu\nu}=\eta^{\mu\nu}-G'^{\mu\nu}$, and
\beqa
G'^{\mu\nu}= \prt_a X^{\mu}  \prt_b X^{\nu} G'^{ab}&;&\Theta^{\mu\nu}= \prt_a X^{\mu}  \prt_b X^{\nu} \Theta^{ab}\labell{G2}
\eeqa
The symmetric matrix $G'^{ab}$ and the antisymmetric matrix $\Theta^{ab}$ are
\beqa
G'^{ab}= \left(\frac{1}{\teta+F} \teta\frac{1}{\teta-F}\right)^{ab}&;&\Theta^{ab} =\left(\frac{1}{\teta+F} F\frac{1}{\teta-F}\right)^{ab} \labell{theta}
\eeqa
In fact it has been observed in \cite{Garousi:2015qgr} that the matrices $G'^{\mu\nu},\, \Theta^{\mu\nu},\, \bot'^{\mu\nu}$ transform among themselves under the T-duality transformation \reef{A'}. When the gauge field strength is zero, $\Theta^{\mu\nu}$ is zero and the matrices $G'^{\mu\nu}$ and $ \bot'^{\mu\nu}$ reduce to the projection metrics $\tG^{\mu\nu}$ and $ \bot^{\mu\nu}$, respectively. 

One may use the extension  $F\rightarrow\tB+F$ in the couplings \reef{final} to find the couplings between massless open string fields and the B-field. The B-field, however, must be along the world-volume directions. Moreover, since the total derivative terms and the terms that involve the Bianchi identity, $dF=0$,  have been ignored in \reef{final}, the above replacement can not correctly produce the couplings involving $dB=H$. In fact, if one uses the above replacement and considers the couplings which have two B-fields, the result would be $\alpha'T_pH_{abc}H^{abc}/6$ up to a total derivative term. The coefficient of this term is not the one in \reef{dbi2} which is consistent with the S-matrix element.
 
However,   the extension  $F\rightarrow\tB+F$ in the couplings \reef{final} produces the correct couplings of massless open string fields in the presence of constant B-field. One may also use  the SW map \cite{Seiberg:1999vs} to transform the couplings to the non-commutative variables. When there is only massless open string fields, one expects that, as in the DBI part   \cite{Seiberg:1999vs},   the commutative fields transform to non-commutative fields, the symmetric part of matrix $1/(\eta+B)$ appears as open string metric for contracting the world-volume indices and the antisymmetric part of this matrix appears in the parameter of the non-commutative    star product. So in the presence of constant B-field, the non-commutative form of the couplings  are the same as \reef{final} in which the open string fields are non-commutative fields, the multiplication rule is $*$-product, and the world-volume indices are contracted with the open string metric. If the couplings \reef{final} are reproduced by S-matrix element in flat space, then  their corresponding non-commutative couplings are reproduced by the S-matrix elements in the presence of constant B-field.  When there are both massless open and closed strings, however, the transformation of world-volume couplings from commutative fields to non-commutative fields is not so easy. One is required to introduce new multiplication rules, \ie $*_n$-product \cite{Garousi:1999ch,Liu:2000ad}. The couplings involving only massless closed string fields are invariant under the SW map.

The Riemann curvature $R$, $\nabla H$ and $\nabla\nabla\phi$ couplings at order $\alpha'^2$ in the world-volume of D-branes in the superstring theory are given in \reef{RTN}. These couplings are invariant under linear T-duality transformations \reef{linear} and are covariant, \ie the   metric contracting the indices, are either the spacetime metric or the first fundamental form. So one can not extend these covariant couplings to include $F_{ab}$ and $\prt_a\chi^i$ by extending the first fundamental form  to $G'^{\mu\nu}$ in which the spacetime metric must be flat. However, if one considers only two NSNS couplings, then the metric for contracting the indices is flat metric. In that case, one may extend the first fundamental form  to \reef{G2} to include $F_{ab}$ and $\prt_a\chi^i$ . In fact, these matrices have been used in \cite{Ghodsi:2016qey} to construct a non-covariant form of the couplings of two NSNS fields at order $\alpha'^2$  in the presence of constant field strength $F_{ab}$ and constant velocity $\prt_a\chi^i$  by requiring the couplings to be invariant under the linear T-duality transformations \reef{linear} and \reef{A'}.   Using the extension $F\rightarrow\tB+F$, then the couplings have been considered for zero $F_{ab}, \prt_a\chi^i$. It has been shown in \cite{Ghodsi:2016qey} that the two NSNS couplings are fully consistent with the $\alpha'^2$-order contact terms of the disk-level S-matrix element of two NSNS vertex operators in the presence of constant B-field in the superstring theory \cite{Garousi:1998bj}. 

\section{Discussion}

In this article, we have reviewed the duality method for finding higher derivative corrections to the supergravities and to the DBI/WZ action. We have seen that to impose the T-duality constraint on the effective actions there are two approaches. One approach is the  covariant approach in which the T-duality invariant action would be  covariant but the T-duality transformations are the Buscher rules plus their  higher derivative corrections. In the non-covariant approach, the T-duality invariant action would be  non-covariant but the T-duality transformations are  the standard Buscher rules. The two T-duality invariant actions should be related by some non-covariant field redefinitions. 

In the covariant approach, we have seen that the T-duality constraint can fix the   O$_p$-plane action in the bosonic string theory at order $\alpha'$, \ie action \reef{boso}, and can fix  the  NSNS couplings of O$_p$-plane action in the type II superstring theory at order $\alpha'^2$, \ie actions \reef{RTN} and \reef{finalH}.  The T-duality transformations that have been used in the bosonic string theory is the standard Buscher rules plus their  derivative corrections at order $\alpha'$, \ie equation \reef{corrT}, whereas, the T-duality transformations that have been used in the superstring theory is only the Buscher rule. This steams from the fact that the first higher derivative corrections to the supergravity is at 8-derivative level. So the higher derivative  corrections to the Buscher rules in the covariant approach  in the superstring theory starts at order $\alpha'^3$. Since the T-duality transformations that have been used in the O$_p$-plane action are the same as the T-duality transformations that have been found from the bulk actions, the above calculations  indicate also that the form of T-duality transformations for bulk actions and for brane actions are identical. 

In the non-covariant approach, we have seen that the T-duality constraint can fix the D$_p$-brane couplings of  massless open string fields at order $\alpha'$ which includes all orders of $F$ and the D$_p$-brane velocity that in our convention are dimensionless, \ie action \reef{final}. The T-duality transformation that has been used is the standard transformation \reef{A'} without $\alpha'$-corrections. Since the metric in contracting the indices in the action \reef{final} is $\eta_{ab}, \, \eta_{ij}$ and the derivatives in this action are also partial  derivatives, the action is not covariant. In a covariant action, the indices would be contracted with the pull-back metric and the derivatives are also covariant derivatives constructed from the pull-back metric. It would be interesting to find such covariant action at order $\alpha'$. This may be done by finding appropriate non-covariant field redefinitions to convert the non-covariant action \reef{final} to the covariant form, or one may find $\alpha'$-corrections to the T-duality transformation \reef{A'} and then find an action at order $\alpha'$ which would be  invariant under such T-duality transformations.

In the covariant approach, the T-duality transformations at the second order of fields are also used to constrain the massless closed string couplings at order $\alpha'$ on the world-volume of D$_p$-brane in the bosonic string theory. This fixes all covariant couplings   up to the terms that have B-field potential, \ie action \reef{dbi2}. It would be interesting to use the full non-linear T-duality transformations to find the contribution of the B-field potential into the covariant action \reef{dbi2} as well. Then the extension $\tB\rightarrow \tB+F$ may be used to find all gauge field couplings at order $\alpha'$ in a covariant action. On the other hand, the covariant form of the transverse scalar fields at order $\alpha'$ are known from the second fundamental forms in  \reef{dbi2}. Making these two covariant couplings to be consistent with T-duality may fix the $\alpha'$-correction to the T-duality transformation \reef{A'}. 

Unlike the T-duality transformations of the massless closed string fields in the superstring theory which receive no corrections at orders $\alpha'$ and $\alpha'^2$, the T-duality transformations for the massless open string fields should receive corrections at order $\alpha'$ as in the bosonic case. Since we do not know the $\alpha'$ corrections to the transformation \reef{A'}, one may use the non-covariant approach to find  the couplings of the massless open string fields   at order $\alpha'^2$ on the world-volume of D$_p$-brane in the superstring theory, \ie analogue of the couplings \reef{final} at order $\alpha'^2$. To perform this calculation, one should consider all possible contractions of $\prt F,\, \prt\prt F,\,\prt\prt\chi,\,\prt\prt\prt\chi$ with matrices $\Theta,\,G',\,\bot'$ at order $\alpha'^2$ with unknown coefficients, and then find the coefficients by constraining them to be invariant under the T-duality transformation \reef{A'}. Such couplings at the level of four fields have been found in \cite{Garousi:2015mdg}.

The T-duality transformations for the massless closed string fields have no corrections at order $\alpha'$ and $\alpha'^2$ in the superstring theory, as a result, the  D$_p$-brane world-volume couplings of  massless closed string fields      at order $\alpha'^2$, which are invariant under the Buscher rules, must be covariant. Such effective action must include the couplings  \reef{finalH} as well as some couplings in which $\nabla\cdots\nabla B$ have even number of transverse indices, and $\nabla\cdots\nabla \phi$, $\nabla\cdots\nabla  R$ have odd number of transverse indices. The difficult part of the calculation is that there may be arbitrary number of B-fields   in each term. So one may consider the simple case that there are only four fields to find the  terms that have only $dB=H$.  The higher order terms, however,  would involve the B-field potential. Such terms then would produce covariant couplings for gauge field $F$ upon replacing  $\tB\rightarrow \tB+F$. The consistency of  such couplings and the covariant transverse scalar couplings in \reef{RTN} with T-dualiy,   may shed light on the  $\alpha'^2$-corrections to the T-duality transformation \reef{A'}.

The $\alpha'$-corrections to the Buscher rules \reef{corrT} have been found in \cite{Kaloper:1997ux} by requiring the known curvature squared corrections to the Einstein gravity \cite{Meissner:1996sa} to be invariant under the T-duality. One may impose the invariance of the effective action at each order of $\alpha'$ under T-duality to find the couplings as well as the covariant corrections to the Buscher rules. We have done this calculation in the bosonic and in the heterotic string theories for the bulk couplings at orders $\alpha'$ and $\alpha'^2$ and for the simple case that the metric is diagonal and B-field is zero. We have found  positive answer \cite{GR}. Such calculations for non-zero B-field at order $\alpha'^2$ and $\alpha'^3$ would  produce all H-couplings that are not known from other approaches in finding the higher derivative couplings  in the string theory.

{\bf Acknowledgments}:   This work is supported by Ferdowsi University of Mashhad. 

 
\end{document}